\documentclass[aps,prd,superscriptaddress,preprintnumbers]{revtex4-2}

\usepackage{amsmath}
\usepackage{amssymb}
\usepackage{graphicx}
\usepackage{hyperref}
\usepackage{slashed}
\usepackage{array}
\usepackage{makecell}
\usepackage{booktabs}
\usepackage{multirow}
\usepackage{subcaption}
\usepackage{tikz}
\usetikzlibrary{arrows.meta}
\usetikzlibrary{decorations.pathmorphing}
\newcommand{\KUST}{Faculty of Science, Kunming University of Science and Technology, Kunming 650500, Yunnan, People's Republic of China}
\newcommand{\SKLP}{State Key Laboratory of Particle Detection and Electronics, University of Science and Technology of China, Hefei 230026, Anhui, People's Republic of China}
\newcommand{\USTC}{Department of Modern Physics, University of Science and Technology of China, Hefei 230026, Anhui, People's Republic of China}
\newcommand{\CCKUST}{City College, Kunming University of Science and Technology, Kunming 650500, Yunnan, People's Republic of China}

\begin{document}
	
	\title{Two-loop QCD corrections to $CP$-even Higgs boson decays into $V\gamma$ in the Type-I Two-Higgs-Doublet Model}
	\thanks{Supported by the National Natural Science Foundation of China (Grant No.~12065013)}

	\date{\today}
	Submitted to Chinese Physics C
	
	\begin{abstract}
		We investigate the QCD corrections to the loop-induced decays of the heavy $CP$-even Higgs boson, $H \to V\gamma$ ($V = Z, \gamma$), within the framework of the Type-I Two-Higgs-Doublet Model (THDM). Owing to the similarity of the underlying loop topologies, the two decay channels exhibit closely correlated behaviour. After imposing relevant phenomenological constraints, we evaluate the decay widths for 5000 viable parameter points at next-to-leading order (NLO) in QCD under the alignment limit. The $H\to\gamma\gamma$ width exceeds that of $H\to Z\gamma$ by roughly a factor of four. Both are dominated by the top-quark loop: the $W$ contribution vanishes in the alignment limit, and the charged-Higgs loop is suppressed throughout most of the viable parameter region. The corrections exhibit a pronounced dependence on the heavy Higgs boson mass $m_H$, attaining a maximum in the vicinity of the $t\bar{t}$ production threshold. To elucidate the individual parameter dependences, dedicated benchmark scenarios are analysed, revealing that the decay widths diminish monotonically with increasing $\tan\beta$ while growing with $m_H$. The QCD corrections for both decay modes typically span approximately $-20\%$ to $+20\%$, with larger corrections in the non-alignment scenario: in the $H\to\gamma\gamma$ channel they reach up to $+38\%$ at large $\tan\beta$, and in the $H\to Z\gamma$ channel they cross zero to become positive. The $CP$ selection rule forces the $t\bar{t}$ pair into a $P$-wave configuration at threshold, which strongly suppresses the Coulomb corrections and secures the perturbative stability of the fixed-order prediction. A comparison with ATLAS limits shows that current data do not yet constrain the parameter space, whereas the $\mathcal{O}(15\%)$ NLO corrections may become relevant for interpreting exclusion bounds at the HL-LHC.
		\begin{description}
			\item[keywords]
			Two-Higgs-Doublet Model, Higgs decays, QCD corrections
		\end{description} 
	\end{abstract}

	\author{Lin-Qiao Qu}
	\affiliation{\KUST}
	
	\author{Zhong-Yuan Liu}
	\affiliation{\SKLP}
	\affiliation{\USTC}
	
	\author{Peng-Fei Duan}
	\email{dpffqs@mail.ustc.edu.cn}
	\affiliation{\KUST}
	\affiliation{\CCKUST}
	
	\author{Yu Zhang}
	\affiliation{\KUST}
	
	\author{Wei-Long Duan}
	\affiliation{\CCKUST}
	
	\maketitle
	
	\allowdisplaybreaks

	\section{Introduction}
	
	\par
	In July 2012, the ATLAS and CMS collaborations independently reported the observation of a new scalar boson at the Large Hadron Collider (LHC)~\cite{ATLAS:2012yve,CMS:2012qbp}. The observed particle was subsequently identified as the Higgs boson predicted by the Standard Model (SM) through comprehensive measurements of its mass, spin, and charge-parity ($CP$) properties~\cite{ATLAS:2015zhl,CMS:2014nkk}. The Higgs boson plays a central role in the electroweak (EW) symmetry breaking mechanism and serves as a sensitive probe of physics beyond the SM (BSM). In particular, loop-induced decay channels such as $h\to V\gamma$ ($V = Z,\, \gamma$) are especially susceptible to virtual contributions from new particles, rendering them powerful windows into potential BSM dynamics.
	
	The $h\to\gamma\gamma$ channel offers excellent mass resolution and was instrumental in the initial discovery. A combined fit incorporating $h \to ZZ^{*} \to 4\ell$ data yields a determination of the Higgs boson mass with a relative precision of 0.09\%~\cite{ATLAS:2023oaq}. Notably, localised excesses have been reported in the diphoton invariant mass spectrum: a local significance of $2.9\sigma$ near $m_{\gamma\gamma}\approx 95~\mathrm{GeV}$~\cite{CMS:2024yhz} and $3.3\sigma$ near $m_{\gamma\gamma}\approx 684~\mathrm{GeV}$~\cite{ATLAS:2021uiz}, which may hint at resonant production of as-yet-unidentified states. The $h\to Z\gamma$ decay, for which first evidence was recently reported at the LHC~\cite{ATLAS:2023yqk}, yields a measured signal strength of $\mu = 2.2 \pm 0.7$ relative to the SM prediction. These intriguing deviations, though not yet statistically conclusive, motivate both improved experimental measurements and refined theoretical predictions within and beyond the SM.
	
	At the high-luminosity phase of the LHC (HL-LHC), the branching ratios of $h\to\gamma\gamma$ and $h\to Z\gamma$ are projected to be measured with anticipated precisions of approximately 1.8\% and 9.8\%, respectively~\cite{Cepeda:2019klc}. Disentangling potential BSM effects at this level of precision necessitates theoretical predictions of commensurate accuracy. In the SM, the $h\to\gamma\gamma$ amplitude was first evaluated at leading order (LO) in Ref.~\cite{Ellis:1975ap}. The next-to-leading order (NLO) QCD corrections were subsequently derived in Refs.~\cite{Zheng:1990qa,Djouadi:1990aj,Dawson:1992cy,Melnikov:1993tj,Inoue:1994jq,Fleischer:2004vb}, whereas NLO EW corrections were incorporated in Refs.~\cite{Aglietti:2004nj,Fugel:2004ug,Degrassi:2005mc,Passarino:2007fp}. The decay width has further been evaluated up to NNLO accuracy~\cite{Steinhauser:1996wy,Maierhofer:2012vv}, with $\text{N}^3\text{LO}$ and $\text{N}^4\text{LO}$ QCD corrections having additionally been determined in Ref.~\cite{Sturm:2014nva}. The $h\to Z\gamma$ decay has been computed to NLO precision, encompassing both QCD~\cite{Spira:1991tj,Bonciani:2015eua,Gehrmann:2015dua} and EW corrections~\cite{Sang:2024vqk,Chen:2024vyn}. Beyond the SM, the process $\phi \to \gamma\gamma$ ($\phi$ denoting a generic neutral Higgs state) has been analysed at NLO within several extended frameworks~\cite{Djouadi:1993ji,Harlander:2005rq,Aglietti:2006tp,Bagnaschi:2022dqz,Degrassi:2023eii}. A number of studies have also investigated BSM effects at LO and examined the correlations between the $\gamma\gamma$ and $Z\gamma$ channels~\cite{Chiang:2012qz,Chen:2013dh,BhupalDev:2013xol,Arhrib:2014pva,Kanemura:2018esc,Benbrik:2022bol}. 
	
	The Two-Higgs-Doublet model (THDM) augments the Higgs sector by an additional scalar doublet, thereby giving rise to five physical Higgs bosons: two $CP$-even ($h$ and $H$), one $CP$-odd ($A$), and two charged bosons ($H^{\pm}$). We identify $h$ with the 125~GeV Higgs boson discovered at the LHC, with $H$ designating the heavier $CP$-even state. In the present work, we investigate the decays $H \to V\gamma$ ($V = Z,\, \gamma$) within the Type-I THDM. The Type-I THDM is generally less constrained by flavour observables, leaving open a broader region of parameter space at low $\tan\beta$.
	The decay of the heavy $CP$-even Higgs boson has been examined at NLO in Ref.~\cite{Kanemura:2022ldq}, wherein the authors utilised the package \texttt{H-COUP}~3.0~\cite{Kanemura:2019slf,Aiko:2023xui}. Nonetheless, their analysis did not encompass the $H \to V\gamma$ channels, and the QCD corrections therein were derived under the large-top-quark-mass approximation, thereby restricting the validity of the results to the regime of relatively light Higgs boson masses. Here, we evaluate the QCD contributions to these processes retaining the full mass dependence and analyse the impact of the relevant THDM parameters.
	
	The remainder of this paper is organised as follows. In Sec.~\ref{sec:model}, we briefly review the THDM and delineate the phenomenological constraints imposed. Sec.~\ref{sec:processes} describes the tensor decomposition of the $H \to V\gamma$ amplitude, the renormalisation procedure, and the computational tools employed. In Sec.~\ref{sec:res}, we present numerical results for the LO and NLO decay widths, analyse their dependence on the THDM parameters, and compare the two channels. Finally, Sec.~\ref{sec:summary} summarises our main results.

	\section{The Two-Higgs-Doublet model \label{sec:model}}
	
	The THDM extends the SM Higgs sector by incorporating two $SU(2)_L$ scalar doublets, $\Phi_1$ and $\Phi_2$, which may be parameterised as
	\begin{equation}
		\Phi_{1}= \begin{pmatrix} \omega_{1}^{+} \\ \dfrac{1}{\sqrt{2}}\left(v_{1}+\xi_{1}+i \chi_{1}\right) \end{pmatrix}, \quad
		\Phi_{2}= \begin{pmatrix} \omega_{2}^{+} \\ \dfrac{1}{\sqrt{2}}\left(v_{2}+\xi_{2}+i \chi_{2}\right) \end{pmatrix}.
	\end{equation}
	Here, $v_i$ denotes the vacuum expectation value (VEV) of $\Phi_i$, with the SM VEV given by $v=\sqrt{v_1^2+v_2^2}=(\sqrt{2}G_F)^{-1/2}\approx 246~\mathrm{GeV}$.
	
	To suppress flavour-changing neutral currents (FCNCs) at tree level, a softly broken discrete $Z_2$ symmetry is imposed, under which $\Phi_1 \to -\Phi_1$. The resulting scalar potential takes the form
	\begin{equation} \label{eq:potential}
		\begin{aligned}
			V =&~ m_{11}^{2} \Phi_{1}^{\dagger} \Phi_{1}  + m_{22}^{2} \Phi_{2}^{\dagger} \Phi_{2}  -  \left[ m_{12}^{2}\Phi_{1}^{\dagger} \Phi_{2}+\mathrm{h.c.}\right]  + \frac{1}{2} \lambda_{1}(\Phi_{1}^{\dagger} \Phi_{1})^{2}  + \frac{1}{2} \lambda_{2}(\Phi_{2}^{\dagger} \Phi_{2})^{2} \\
			& + \lambda_{3}(\Phi_{1}^{\dagger} \Phi_{1})(\Phi_{2}^{\dagger} \Phi_{2})  + \lambda_{4}(\Phi_{1}^{\dagger} \Phi_{2})(\Phi_{2}^{\dagger} \Phi_{1})  + \frac{1}{2} \left[\lambda_{5}(\Phi_{1}^{\dagger} \Phi_{2})^{2}+\mathrm{h.c.}\right].
		\end{aligned}
	\end{equation}
	This potential involves five dimensionless couplings $\lambda_i$ and three mass-dimension-two parameters $m_{ij}^2$. We restrict our attention to the $CP$-conserving scenario, in which all parameters are taken to be real.
	
	The mass eigenstates are obtained by diagonalising the quadratic terms of the scalar potential and are related to the gauge eigenstates via
	\begin{equation}
		\begin{aligned}
			\left(\!\begin{array}{c} H \\ h \end{array}\!\right) &=
			\mathbf{R}^{\top}(\alpha)
			\left(\!\begin{array}{c} \xi_{1} \\ \xi_{2} \end{array}\!\right), \quad
			\left(\!\begin{array}{c} G^{0} \\ A \end{array}\!\right) =
			\mathbf{R}^{\top}(\beta)
			\left(\!\begin{array}{c} \chi_{1} \\ \chi_{2} \end{array}\!\right), \quad
			\text{and} \quad
			\left(\!\begin{array}{c} G^{\pm} \\ H^{\pm} \end{array}\!\right) =
			\mathbf{R}^{\top}(\beta)
			\left(\!\begin{array}{c} \omega_{1}^{\pm} \\ \omega_{2}^{\pm} \end{array}\!\right).
		\end{aligned}
	\end{equation}
	Here,
	\begin{equation}
		\mathbf{R}(\theta) =
		\begin{pmatrix}
			c_{\theta} & -s_{\theta} \\
			s_{\theta} &  c_{\theta}
		\end{pmatrix},
		\qquad
		c_{\theta} \equiv \cos\theta, \quad
		s_{\theta} \equiv \sin\theta.
	\end{equation}
	Here $\alpha$ and $\beta$ denote the mixing angles. The ratio of the two VEVs is defined as $\tan\beta = v_{2}/v_{1}$, and the mixing angle $\alpha$ is fixed by the relation
	\begin{equation}
		\tan 2\alpha = \frac{\left[M^{2}-(\lambda_{3}+\lambda_{4}+\lambda_{5})v^{2}\right]s_{2\beta}}
		{c_{\beta}^{2}(M^{2}-\lambda_{1}v^{2})-s_{\beta}^{2}(M^{2}-\lambda_{2}v^{2})},
	\end{equation}
	where $M^{2} \equiv m_{12}^{2}/(s_{\beta}c_{\beta})$. Upon spontaneous symmetry breaking, the physical scalar spectrum comprises two $CP$-even Higgs bosons ($h$, $H$), one $CP$-odd Higgs boson ($A$), and a pair of charged Higgs bosons ($H^{\pm}$), with the three Goldstone bosons ($G^{0}$, $G^{\pm}$) being absorbed as the longitudinal degrees of freedom of the $Z$ and $W^{\pm}$ gauge bosons. The masses of the physical Higgs bosons are given by
	\begin{equation}
		\begin{aligned}
			& m_{H,h}^2=\frac{1}{2}\left\{ M^2+v^2(\lambda_1 c^2_{\beta} +\lambda_2 s^2_{\beta}) \right. \mp \frac{1}{c_{2\alpha} }\left. \left[M^2 c_{ 2\beta}-v^2(\lambda_1 c^2_{\beta} -\lambda_2 s^2_{\beta}) \right] \right\} , \\
			& m_{H^{\pm}}^2=M^2 -\frac{1}{2} \left(\lambda _4+\lambda _5\right) v^2 , \\
			& m_{A}^2=M^2-\lambda _5 v^2 .
		\end{aligned}
	\end{equation}
	
	The four types of THDM are distinguished by the assignment of Yukawa couplings between the two Higgs doublets and the right-handed fermions~\cite{Barger:1989fj, Arco:2022xum}. The Yukawa Lagrangian reads
	\begin{equation}
		\begin{aligned}
			\mathcal{L}_{\mathrm{Yuk}}^{\mathrm{THDM}} =& -\sum _{f,k}\frac{m_{f}}{v}\Bigl[\bar{f} f (h\kappa _{f,h} +H\kappa _{f,H} )-2iI_{f}^3\bar{f} \gamma _{5} f A\kappa _{f,A}\Bigr]\\
			& +\frac{\sqrt{2}}{v}\Bigl[\sum _{i,j} V_{CKM}^{ij}\bar{u}_{i}\left(m_{u}^{i} \kappa _{u,A} \mathcal{P}_{L} -m_{d}^{j} \kappa _{d,A}\mathcal{P}_{R}\right) d_{j} H^{+} -\sum _{i}\bar{\nu }_{i} m_{l}^{i} P_{R} l_{i} H^{+} \kappa _{l,A} +\mathrm{h.c.}\Bigr],
		\end{aligned}
	\end{equation}
	where $I_f^3 = \pm \tfrac{1}{2}$ denotes the third component of the weak isospin, $u_i$ and $d_i$ represent the up- and down-type quarks, $\nu_i$ and $l_i$ denote the neutrinos and charged leptons, $\mathcal{P}_{L/R}$ are the chiral projection operators, and $V_{CKM}^{ij}$ is the Cabibbo--Kobayashi--Maskawa (CKM) matrix element. The coupling modifiers $\kappa_{f,\phi}$ satisfy the relations
	\begin{equation}\label{eq:kappa}
		\kappa_{f,h} = s_{\beta-\alpha} + \kappa_{f,A} c_{\beta-\alpha}, \qquad
		\kappa_{f,H} = c_{\beta-\alpha} - \kappa_{f,A} s_{\beta-\alpha},
	\end{equation}
	where the values of $\kappa_{f,A}$ in the four types of THDM are listed in Table~\ref{tab:yukawa}.
	
	\begin{table}[h]
		\centering
		\setlength{\tabcolsep}{5pt}
		\begin{tabular}{c|cccc}
			\hline
			& Type-I & Type-II & Type-III & Type-IV \\
			\hline
			$\kappa_{u,A}$ & $\cot\beta$ & $\cot\beta$ & $\cot\beta$ & $\cot\beta$ \\
			$\kappa_{d,A}$ & $\cot\beta$ & $-\tan\beta$ & $-\tan\beta$ & $\cot\beta$ \\
			$\kappa_{l,A}$ & $\cot\beta$ & $-\tan\beta$ & $\cot\beta$ & $-\tan\beta$ \\
			\hline
		\end{tabular}
		\caption{The values of $\kappa_{f,A}$ in the four types of THDM, where $u$, $d$, and $l$ represent up-type quarks, down-type quarks, and leptons, respectively. \label{tab:yukawa}}
	\end{table}
	
	For the subsequent phenomenological analysis, the Higgs sector is conveniently parameterised by the following eight independent quantities:
	\begin{equation}
		m_h,~ m_H,~ m_A,~ m_{H^{\pm}},~ M^2,~ s_{\beta-\alpha},~ \tan\beta,~ v,
	\end{equation}
	where $m_h \simeq 125~\mathrm{GeV}$ and $v \simeq 246~\mathrm{GeV}$ are fixed at their SM values, and we consider $s_{\beta-\alpha}$ and $\tan\beta \ge 0$.
	
	To ensure the phenomenological viability of the model, we impose both theoretical and experimental constraints on the parameter space. The constraints adopted are enumerated below:
	
	\begin{itemize}
		\item \textit{Vacuum stability}: Vacuum stability requires the scalar potential to be bounded from below, leading to the following conditions on the quartic couplings $\lambda_i$~\cite{Sher:1988mj}:
		\begin{equation}
			\begin{aligned}
				& \lambda_1>0,\quad \lambda_2>0,\quad \lambda_3>-\sqrt{\lambda_1 \lambda_2}, & \lambda_3+\lambda_4-|\lambda_5|>-\sqrt{\lambda_1 \lambda_2}.
			\end{aligned}
		\end{equation}
		
		\item \textit{Perturbativity and unitarity}: Perturbativity is imposed to guarantee the reliability of the perturbative expansion. We require that the quartic scalar couplings remain below the perturbative bound, e.g.\  $|\lambda_i| \leq 4\pi$, or equivalently that the physical quartic couplings satisfy $|g_{h_i h_j h_k h_l}| \leq 4\pi$. Furthermore, tree-level unitarity of scalar--scalar scattering amplitudes places additional constraints on certain linear combinations of the quartic couplings in the scalar potential. The complete set of unitarity conditions in the THDM is collected in~\cite{Ginzburg:2005dt}.
		
		\item \textit{Oblique parameters}: Electroweak precision observables furnish stringent constraints on new physics (NP). The oblique parameters $S$, $T$, and $U$~\cite{Peskin:1990zt,Peskin:1991sw}, defined in terms of gauge-boson self-energies, are particularly sensitive to loop-level manifestations of NP. In the THDM, one-loop contributions to these parameters severely constrain the mass splittings among the non-SM-like Higgs bosons~\cite{Chowdhury:2017aav,Haller:2018nnx,Wang:2022yhm}. To satisfy electroweak precision constraints and to simplify the parameter scan, we assume mass degeneracy among the additional Higgs bosons, i.e.\ $m_H = m_A = m_{H^{\pm}}$. For the oblique parameters, we employ the values extracted from the most recent PDG global electroweak fit~\cite{ParticleDataGroup:2024cfk} as
		\begin{equation}
			\begin{array}{lll}
				S = -0.04 \pm 0.10, & \qquad T = 0.01 \pm 0.12, &\qquad  U = -0.01 \pm 0.09, \\
				\rho_{\mathrm{ST}} = 0.93, & \qquad \rho_{\mathrm{SU}} = -0.70, & \qquad  \rho_{\mathrm{TU}} = -0.87,
			\end{array}
		\end{equation}
		where $\rho_{ij}$ denote the elements of the correlation matrix.
		
		\item \textit{SM-like Higgs coupling strength measurements}: The couplings of the observed 125~GeV Higgs boson have been precisely determined at the LHC~\cite{ATLAS:2022vkf,CMS:2022dwd}. Within the THDM framework, these measurements constrain the couplings of $h$ to the gauge bosons $W$, $Z$, and $\gamma$, as well as to fermions via the modifiers $\kappa_{f,h}$. The parameter $s_{\beta-\alpha}$ is consequently driven toward unity, corresponding to the so-called \textit{alignment limit}~\cite{Chowdhury:2017aav,Haller:2018nnx}.
		
		\item \textit{Direct searches for exotic Higgs bosons}: Although no statistically significant excess attributable to additional Higgs states has been observed~\cite{Tao:2024syy,ATLAS:2024itc}, direct searches have established exclusion bounds on their masses and couplings. For instance, in the Type-I THDM, LEP data constrain $m_{H^{\pm}} > 75$~GeV~\cite{ALEPH:2013htx}, while LHC data require $m_{H^{\pm}} > 150~\mathrm{GeV}$ in the Type-II THDM~\cite{Arbey:2017gmh}. The parameter $\tan\beta$ is subject to more stringent constraints in the Type-II, -III, and -IV scenarios relative to Type-I, owing to measurements of $\phi \to f\bar{f}$ processes in which the couplings to leptons or down-type quarks exhibit a $\tan\beta$-enhanced scaling.
	\end{itemize}
	
	Theoretical predictions and constraints are evaluated using \texttt{2HDMC}~\cite{Harlander:2013qxa,Eriksson:2009ws,Eriksson:2010zzb}. Constraints from SM-like Higgs boson measurements and from searches for exotic Higgs states are enforced via \texttt{HiggsSignals}~\cite{Bechtle:2013xfa,Stal:2013hwa,Bechtle:2014ewa,Bechtle:2020uwn} and \texttt{HiggsBounds}~\cite{Bechtle:2008jh,Bechtle:2011sb,Bechtle:2012lvg,Bechtle:2013wla,Bechtle:2015pma,Bechtle:2020pkv}, respectively. A parameter point is deemed viable provided the theoretical predictions remain consistent with the experimental data at the 95\% confidence level.

	\section{Calculation setup\label{sec:processes}}
	
	In general, the amplitude for $H\to V\gamma$ may be decomposed as
	\begin{equation}
		\mathcal{M} \;=\; \mathcal{M}^{\mu\nu}\, \varepsilon^{*}_{\mu}(p_{2}) \, \varepsilon^{*}_{\nu}(p_{3}) ,
	\end{equation}
	where $\mathcal{M}^{\mu\nu}$ denotes the tensor amplitude, and $\varepsilon _{\mu }^{*} (p_{2} )$ and $\varepsilon _{\nu }^{*} (p_{3} )$ denote the polarisation vectors of $V$ and $\gamma$, respectively. The most general Lorentz decomposition of $\mathcal{M}^{\mu\nu}$ reads
	\begin{equation}
		\mathcal{M}^{\mu \nu } = p_{2}^{\mu } p_{2}^{\nu } A_{1} + p_{3}^{\mu } p_{3}^{\nu } A_{2} + p_{2}^{\mu } p_{3}^{\nu } A_{3} + p_{3}^{\mu } p_{2}^{\nu } A_{4} + g^{\mu \nu } A_{5} + \epsilon ^{\mu \nu \rho \sigma } p_{2\rho } p_{3\sigma } A_{6}.
		\label{amplitude-all}
	\end{equation}
	Imposing the Ward identity, $\mathcal{M}^{\mu\nu}p_{3\nu}=0$, one obtains
	\begin{equation}
		\label{eq:ward}
		\begin{aligned}
			A_{1} =0,&\quad( p_{2} \cdot p_{3}) A_{4} +A_{5}=0.
		\end{aligned}
	\end{equation}
	For $C$-conserving interactions, $A_6$ vanishes. Furthermore, by virtue of the transversality condition $p_{i}^{\nu}\cdot \varepsilon^*_{\nu}(p_i)=0$, the coefficients $A_2$ and $A_3$ do not contribute. Thus,
	\begin{equation}
		\begin{aligned}
			\mathcal{M}^{\mu\nu}
			&= p_3^{\mu} p_2^{\nu} A_4 + g^{\mu\nu} A_5 \\
			&= \left( p_3^{\mu} p_2^{\nu} - p_2 \!\cdot\! p_3\, g^{\mu\nu} \right) A_4\,.
			\label{eq:amp}
		\end{aligned}
	\end{equation}
	The partial decay widths are accordingly given by
	\begin{equation}
		\Gamma = \begin{cases}
			\dfrac{m_{H}^{6}}{64\pi m_{H}^{3}} |A_{4} |^{2}, & H\to \gamma\gamma\\[6pt]
			\dfrac{\left( m_{H}^{2} - m_{Z}^{2}\right)^{3}}{32\pi m_{H}^{3}} |A_{4} |^{2}, & H\to Z\gamma
		\end{cases}\,.
		\label{eq:width}
	\end{equation}
	Note that the $H\to\gamma\gamma$ width includes the statistical factor of $1/2$ for identical final-state photons.
	
\begin{figure}
		\centering
		\begin{subfigure}[b]{\textwidth}
			\centering
			\includegraphics[scale=0.8]{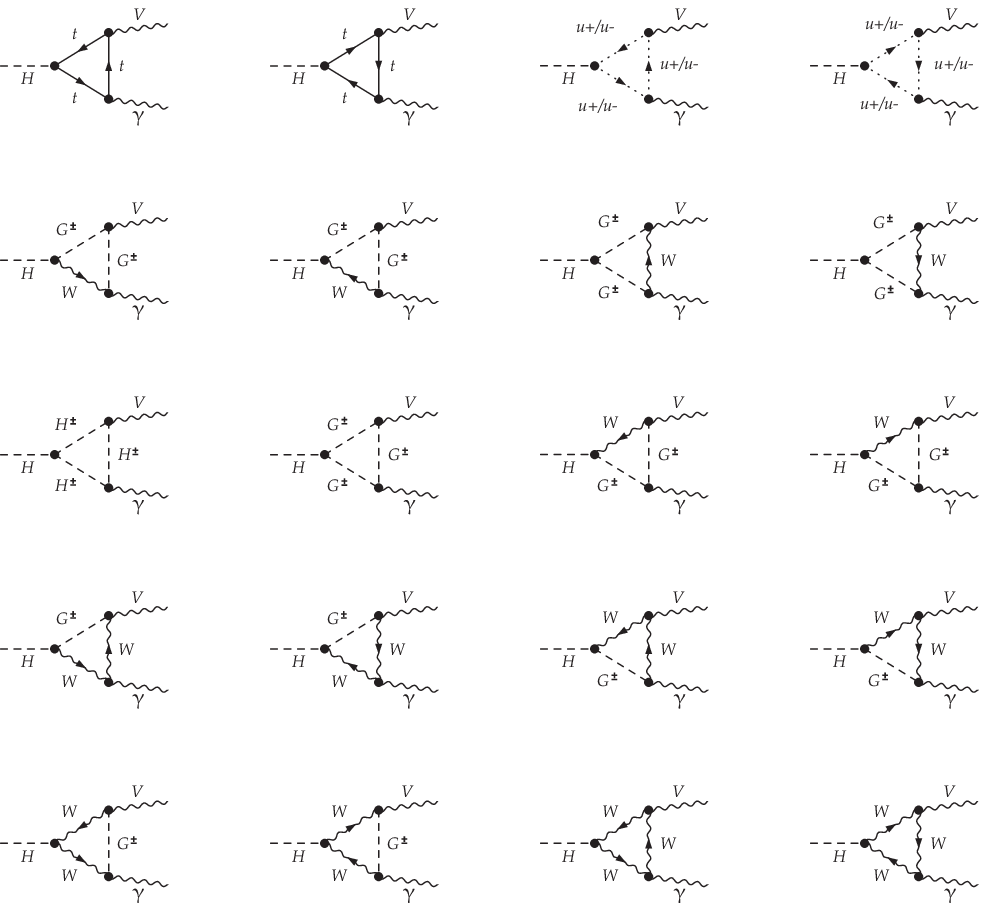}
			\caption{Triangle diagrams for the $H \to V\gamma ~(V=Z,\gamma)$ in the THDM at LO \label{fig:feynman1}}
		\end{subfigure} \\[4pt]
		\begin{subfigure}[b]{\textwidth}
			\centering
			\includegraphics[scale=0.8]{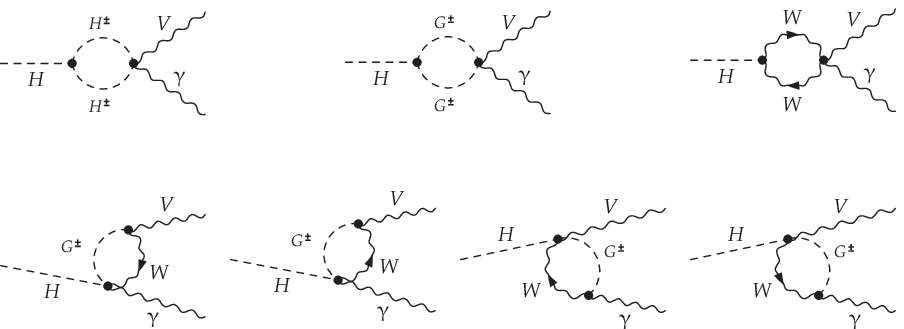}
			\caption{Diagrams with quartic couplings for the $H \to V\gamma ~(V=Z,\gamma)$ in the THDM at LO \label{fig:feynman2}}
		\end{subfigure} \\[4pt]
		\begin{subfigure}[b]{\textwidth}
			\centering
			\includegraphics[scale=0.8]{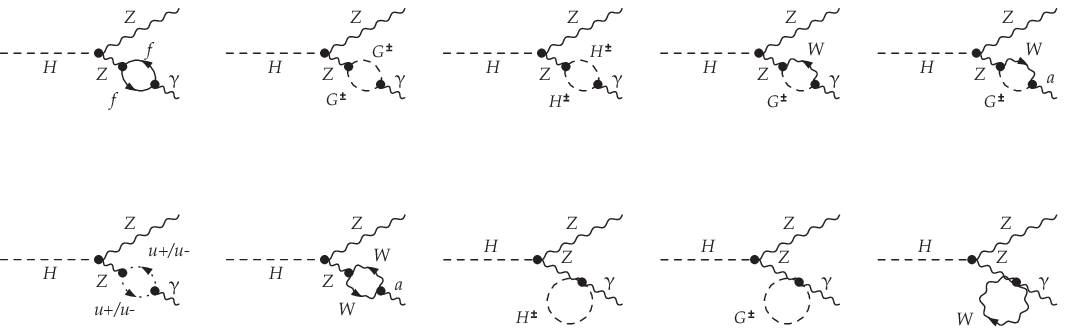}
			\caption{Self-energy diagrams for the $H \to Z\gamma$ process in the THDM at LO \label{fig:feynman3}}
		\end{subfigure}
		\caption{Feynman diagrams for the $H \to V\gamma ~(V=Z,\gamma)$ process in the THDM at LO \label{fig:LO}}
	\end{figure}

	Both processes are loop-induced. The corresponding LO Feynman diagrams are depicted in Figure~\ref{fig:LO}. At LO, 31 and 50 Feynman diagrams contribute to the $H \to \gamma\gamma$ and $H \to Z\gamma$ processes, respectively. The topologies common to both processes are displayed in Figures~\ref{fig:feynman1} and~\ref{fig:feynman2}. The self-energy diagrams arise exclusively in the $H\to Z\gamma$ channel, as illustrated in Fig.~\ref{fig:feynman3}. These diagrams contribute solely to $A_2$ and $A_5$ and serve to cancel ultraviolet divergences. In practice, if one first applies Eq.~\eqref{eq:ward} and retains only the form factor $A_4$, the self-energy diagrams may be discarded directly. We have explicitly verified the Ward identity at both LO and NLO.
	The LO amplitude admits a decomposition into three contributions: the top-quark loops, the $W$-boson sector (diagrams involving the $W$ boson, ghosts, or Goldstone bosons), and the charged-Higgs loops. These combine as
	\begin{equation}
		A_4^{\text{(1-loop)}} = A_{\mathrm{top}}^{\text{(1-loop)}} + A_{W}^{\text{(1-loop)}} + A_{H^\pm}^{\text{(1-loop)}}.
	\end{equation}
	The corresponding LO expressions are collected in Appendix~\ref{sec:LOexpress}, and the relevant THDM couplings for both processes are summarised in Table~\ref{tab:vertices}.
	\begin{table}[htbp]
		\centering
		\renewcommand{\arraystretch}{1.8}
		\setlength{\tabcolsep}{32pt}
		\begin{tabular}{l}
			\hline
			\multicolumn{1}{c}{\textbf{(I) Common vertices for $H\to\gamma\gamma$ and $H\to Z\gamma$}} \\
			\hline
			$g_{H t\bar{t}} = -i\dfrac{m_t}{v}\kappa_{u,H}$ \\[6pt]
			$g_{H G^{+}G^{-}} = i\dfrac{m_H^2}{v}c_{\beta-\alpha}$ \\[6pt]
			$g_{H W^{+}G^{-}} = i\dfrac{m_W}{v}c_{\beta-\alpha}$ \\[6pt]
			$g_{H W^{+}W^{-}} = -i\dfrac{2m_W^2}{v}c_{\beta-\alpha}$ \\[6pt]
			$g_{H u_{+}\bar{u}_{+}} = -i\dfrac{m_W^2}{v}c_{\beta-\alpha}$ \\[6pt]
			$g_{\gamma H^{-}H^{+}} = -ie$ \\[6pt]
			$g_{\gamma\gamma H^{-}H^{+}} = -2ie^2$ \\[6pt]
			$g_{H H^{+}H^{-}} = -\dfrac{i}{v}\Bigl[2(m_H^2 - M^2)\cot2\beta s_{\beta-\alpha}  + (2M^2 - 2m_{H^{\pm}}^2 - m_H^2)c_{\beta-\alpha}\Bigr]$ \\[6pt]
			$g_{H\gamma W^{+} G^{-}} = i\dfrac{e^2}{2s_W}c_{\beta-\alpha}$ \\[6pt]
			\hline
			\multicolumn{1}{c}{\textbf{(II) Vertices unique to $H\to Z\gamma$}} \\
			\hline
			$g_{H Z W^{+} G^{-}} = i\dfrac{e^2}{2c_W}c_{\beta-\alpha}$ \\[6pt]
			$g_{Z H^{+}H^{-}} = -\dfrac{ie}{2c_W s_W}(c_W^2 - s_W^2)$ \\[6pt]
			$g_{\gamma Z H^{+}H^{-}} = -\dfrac{ie^2}{c_W s_W}(c_W^2 - s_W^2)$ \\[6pt]
			\hline
		\end{tabular}
		\caption{Relevant couplings for $H \to \gamma\gamma$ and $H \to Z\gamma$ in the Type-I THDM. \label{tab:vertices}}
	\end{table}
	
	\begin{figure*}[p]
		\begin{subfigure}[b]{\textwidth}
			\centering
			\includegraphics[scale=1]{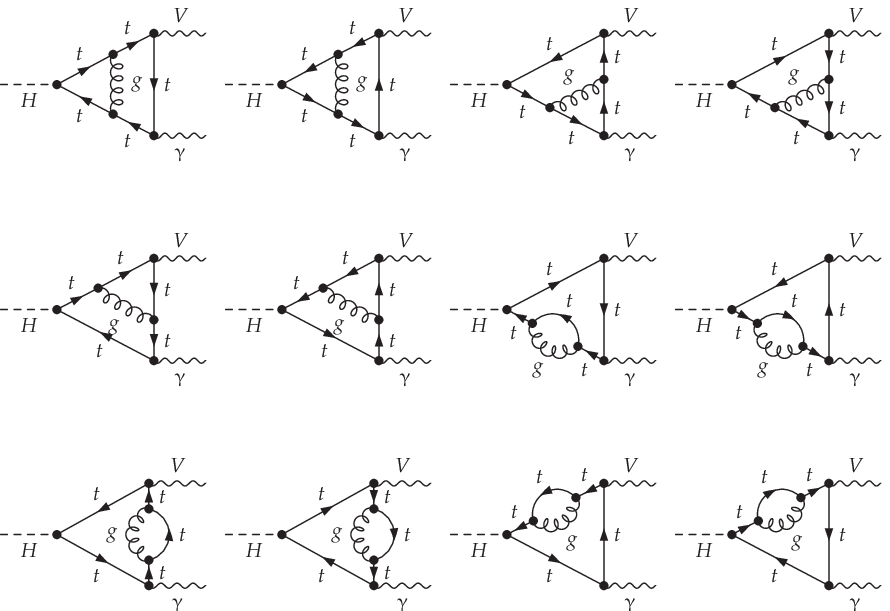}
			\caption{Two-loop Feynman diagrams for the $H \to V\gamma ~(V=Z,\gamma)$ in the THDM at NLO QCD \label{fig:feynman4}}
		\end{subfigure} \\[4pt]
		\begin{subfigure}[b]{\textwidth}
			\centering
			\includegraphics[scale=1]{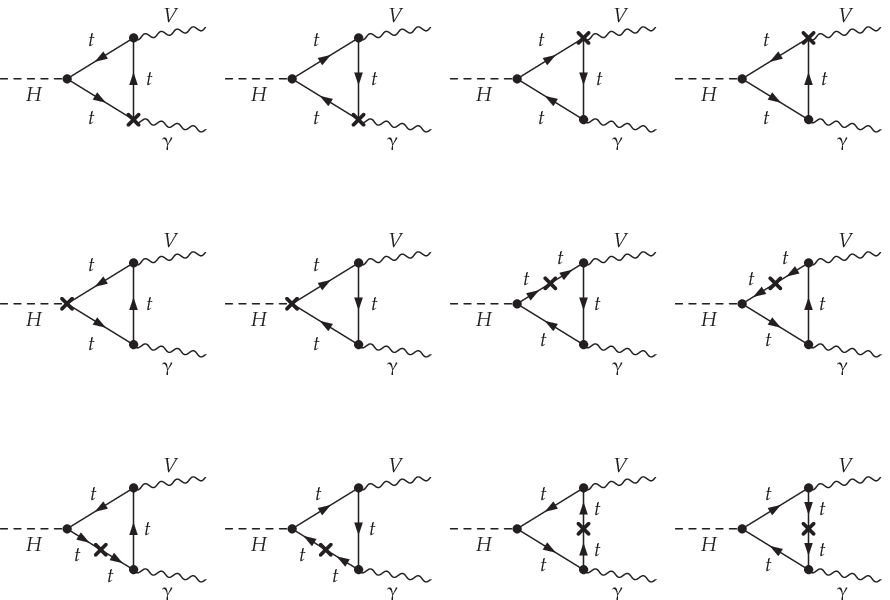}
			\caption{Counterterm diagrams for the $H \to V\gamma ~(V=Z,\gamma)$ in the THDM at NLO QCD \label{fig:feynman5}}
		\end{subfigure}
		\caption{Feynman diagrams for the $H \to V\gamma ~(V=Z,\gamma)$ in the THDM at NLO QCD}
		\label{fig:NLO}
	\end{figure*}
	
	The NLO QCD Feynman diagrams are displayed in Figure~\ref{fig:NLO}. The calculation involves 24 two-loop diagrams together with the associated counterterm diagrams. The counterterm diagrams include the $Z$--$\gamma$ mixing contribution relevant to the $H\to Z\gamma$ channel. In the on-shell (OS) renormalisation scheme, the following identity holds for an on-shell external photon:
	\begin{equation}
		\begin{gathered}
			\begin{tikzpicture}[baseline=-0.6ex,scale=0.8]
				\draw[thick,dashed] (0,0)--(0,1.4);
				\node[left] at (0,1.0) {\small$H$};
				\draw[thick,decorate,decoration={snake,amplitude=1.5pt,segment length=5pt}]
				(0,0)--(-1.2,-0.8);
				\node[left] at (-1.1,-0.8) {\small$Z$};
				\draw[thick,decorate,decoration={snake,amplitude=1.5pt,segment length=5pt}]
				(0,0)--(1.2,-0.8);
				\node[right] at (1.2,-0.8) {\small{$\gamma$}};
				\draw[thick] (-0.15,-0.15)--(0.15,0.15);
				\draw[thick] (-0.15,0.15)--(0.15,-0.15);
			\end{tikzpicture}
		\end{gathered}
		=-
		\begin{gathered}
			\begin{tikzpicture}[baseline=-0.6ex,scale=0.8]
				\draw[thick,dashed] (0,0)--(0,1.4);
				\node[left] at (0,1.0) {\small$H$};
				\draw[thick,decorate,decoration={snake,amplitude=1.5pt,segment length=5pt}]
				(0,0)--(-1.2,-0.8);
				\node[left] at (-1.1,-0.8) {\small$Z$};
				\draw[thick,decorate,decoration={snake,amplitude=1.5pt,segment length=5pt}]
				(0,0)--(1.2,-0.8);
				\node[right] at (1.1,-0.8) {\small$\gamma$};
				\node[left] at (-0.15+0.6,-0.5) {\small$Z$};
				\draw[thick] (-0.15+0.6,-0.15-0.4)--(0.15+0.6,0.15-0.4);
				\draw[thick] (-0.15+0.6,0.15-0.4)--(0.15+0.6,-0.15-0.4);
			\end{tikzpicture}
		\end{gathered}
		=
		\begin{gathered}
			\begin{tikzpicture}[baseline=-0.6ex,scale=0.8]
				\draw[thick,dashed] (0,0)--(0,1.4);
				\node[left] at (0,1.0) {\small$H$};
				\draw[thick,decorate,decoration={snake,amplitude=1.5pt,segment length=5pt}]
				(0,0)--(-1.2,-0.8);
				\node[left] at (-1.1,-0.8) {\small$Z$};
				\draw[thick,decorate,decoration={snake,amplitude=1.5pt,segment length=5pt}]
				(0,0)--(1.2,-0.9);
				\node[right] at (1.2,-0.8) {\small$\gamma$};
				\node[left] at (-0.15+0.6,-0.5) {\small$Z$};
				\fill[color=gray!60] (0.6,-0.4) circle (0.25);
				\draw (0.6,-0.4) circle (0.25);
			\end{tikzpicture}
		\end{gathered}
		\label{eq:Zgamma-identity}
	\end{equation}
	The rightmost diagram corresponds to the self-energy diagram shown in Fig.~\ref{fig:feynman3}. This identity is readily established by writing out the relevant amplitudes explicitly. The contribution of the $HZ\gamma$ vertex counterterm (first diagram) is
	\begin{equation}
		\frac{1}{2}\delta Z_{Z\gamma}\left(i g_{\mu\nu}\frac{2m_Z^2}{v}\right) \epsilon^{\nu*}(p),
	\end{equation}
	while the $Z$--$\gamma$ mixing counterterm insertion (second diagram) evaluates to
	\begin{align}
		\lim_{p^2\to 0} i g_{\mu\rho}\frac{2m_Z^2}{v}
		\frac{-i\left(g^{\rho\sigma}-\frac{p^{\rho}p^{\sigma}}{m_Z^2}\right)}{p^2-m_Z^2}
		(-i g_{\sigma\nu})\left[\frac{1}{2}\bigl(\delta Z_{\gamma Z}+\delta Z_{Z\gamma}\bigr)p^2-\frac{m_Z^2}{2}\delta Z_{Z\gamma}\right]
		\epsilon^{\nu*}(p)
		= -\frac{1}{2}\delta Z_{Z\gamma}\left(i g_{\mu\nu}\frac{2m_Z^2}{v}\right) \epsilon^{\nu*}(p).
	\end{align}
	Thus the first two diagrams cancel exactly, and only the renormalised $Z$--$\gamma$ mixing self-energy insertion on the external photon leg survives. This cancellation, which guarantees that no net $H\to ZZ$ contribution enters the $H\to Z\gamma$ amplitude through the mixing counterterm, is the standard treatment adopted in previous $HZ\gamma$ calculations~\cite{Djouadi:1996ws,Arhrib:2014pva,Sang:2017vph}. Moreover, at NLO QCD, the unrenormalised $Z$--$\gamma$ mixing self-energy vanishes for an on-shell external photon, ${\Sigma}_{Z\gamma}(0)=0$~\cite{Djouadi:1993ss}, so no additional mixing contribution is required at this order. Because all fermion Yukawa couplings in the Type-I THDM scale as $\cot\beta$, the bottom-quark loop is suppressed relative to the top-quark one by $(m_b/m_t)^2\approx 5\times 10^{-4}$ and can be neglected. 
	
	The top-quark wave-function renormalisation constant $\delta Z_f$ does not appear explicitly in the final physical amplitude, as its contributions cancel in the sum of all counterterm diagrams. For a fermion field, the bare and renormalised fields are related by
	\begin{equation}
		f_0 = \bigl(1 + \tfrac12 \delta Z_f\bigr) f,
	\end{equation}
	so that the fermion propagator counterterm reads
	\begin{equation}
		\begin{gathered}
			\begin{tikzpicture}[baseline=-0.6ex, scale=0.8]
				\draw[thick] (-1.2,0) -- (1.2,0);
				\draw[thick] (-0.12,-0.12) -- (0.12,0.12);
				\draw[thick] (-0.12,0.12) -- (0.12,-0.12);
				\draw[thick, -{Stealth[scale=1.2]}] (1.2,0) -- (0.4,0);
				\draw[thick, -{Stealth[scale=1.2]}] (0,0) -- (-0.7,0);
			\end{tikzpicture}
		\end{gathered}
		= i\delta Z_f(\slashed{p}-m).
	\end{equation}
	Inserting this counterterm into the fermion propagator gives
	\begin{equation}
		\begin{gathered}
			\begin{tikzpicture}[baseline=-0.6ex, scale=0.8]
				\fill (-1.2,0) circle (2pt);
				\fill (1.2,0) circle (2pt);
				\draw[thick] (-1.2,0) -- (1.2,0);
				\draw[thick] (-0.12,-0.12) -- (0.12,0.12);
				\draw[thick] (-0.12,0.12) -- (0.12,-0.12);
				\draw[thick, -{Stealth[scale=1.2]}] (1.2,0) -- (0.4,0);
				\draw[thick, -{Stealth[scale=1.2]}] (0,0) -- (-0.7,0);
			\end{tikzpicture}
		\end{gathered}
		= \frac{i}{\slashed{p}-m}\, i\delta Z_f(\slashed{p}-m)\, \frac{i}{\slashed{p}-m}
		= -\delta Z_f \times
		\begin{gathered}
			\begin{tikzpicture}[baseline=-0.6ex, scale=0.8]
				\fill (0.0,0) circle (2pt);
				\fill (1.5,0) circle (2pt);
				\draw[thick] (0,0) -- (1.5,0);
				\draw[thick, -{Stealth[scale=1.2]}] (1.4,0) -- (0.6,0);
			\end{tikzpicture}
		\end{gathered}
	\end{equation}
	For $Sf\bar{f}$ and $Vf\bar{f}$ vertices, similar factors from the fermion-field renormalisation enter the vertex counterterms. Summing over all counterterm diagrams, the contributions from $\delta Z_f$ cancel exactly, leaving no net effect in the physical NLO amplitude.
	The sole renormalisation constant required is the quark-mass counterterm, which in the OS scheme takes the form~\cite{Bernreuther:2004ih}
	\begin{equation}
		\delta m_q = -m_q\frac{\alpha_s}{\pi}\Big(\frac{\mu^2}{m_q^2}\Big)^\epsilon\frac{C_F}{4}\frac{3-2\epsilon}{\epsilon(1-2\epsilon)}(4\pi)^\epsilon\Gamma(1+\epsilon),
	\end{equation}
	where $\mu$ is the renormalisation scale, $C_F = 4/3$, and $d = 4 - 2\epsilon$ is the dimension of spacetime.
	
	In this work, the one-loop form factors are denoted by $A_i^{\text{(1-loop)}}$, whereas the two-loop contributions at $\mathcal{O}(\alpha_s)$ are designated as $A_i^{\text{(2-loop)}}$. At LO, the squared amplitude reduces to
	\begin{equation}
		|A_4|^2 = |A_4^{\text{(1-loop)}}|^2.
	\end{equation}
	At NLO, the squared amplitude is expanded to $\mathcal{O}(\alpha_s)$, yielding
	\begin{equation}
		|A_4|^2 = |A_4^{\text{(1-loop)}}|^{2} + 2\Re[A_4^{\text{(2-loop)}*} A_4^{\text{(1-loop)}}].
	\end{equation}
	The decay width at each perturbative order is then obtained by substituting the corresponding squared amplitude into Eq.~\eqref{eq:width}.
	
	To quantify the impact of QCD corrections beyond LO, we introduce the relative correction defined as
	\begin{equation}
		\delta_{\mathrm{QCD}} \equiv \frac{\Delta_{\mathrm{QCD}}}{\Gamma_{\mathrm{LO}}},
	\end{equation}
	where $\Delta_{\mathrm{QCD}}=\Gamma_{\mathrm{NLO}} - \Gamma_{\mathrm{LO}}$.
	
	The Feynman diagrams and corresponding amplitudes are generated with the \texttt{Mathematica} package \texttt{FeynArts 3.11}~\cite{Hahn:2000kx} and subsequently simplified using \texttt{FeynCalc 10.0.0}~\cite{Mertig:1990an,Shtabovenko:2016sxi,Shtabovenko:2020gxv,Shtabovenko:2023idz}. All calculations are performed in the 't~Hooft--Feynman gauge. At LO, the one-loop integrals are decomposed into Passarino--Veltman functions and numerically evaluated with \texttt{Package-X}~\cite{Patel:2015tea,Patel:2016fam}. The NLO QCD corrections entail a substantial number of two-loop scalar integrals. The integration-by-parts (IBP) reduction is performed with \texttt{Kira 2.3}~\cite{Klappert:2020nbg,Maierhofer:2017gsa}. The resulting master integrals (MIs) are computed using the auxiliary mass flow (AMF) method~\cite{Liu:2017jxz,Liu:2021wks,Liu:2022mfb}, which furnishes an efficient numerical framework based on differential equations with systematically determined boundary conditions. In practice, we utilise the package \texttt{AMFlow}~\cite{Liu:2022chg}, which implements the AMF method, to obtain high-precision numerical results for all MIs. Finally, the one-loop results have been cross-checked against the program \texttt{H-COUP}~3.0, thereby validating the correctness of our implementation.

	\section{Numerical results and discussion\label{sec:res}}
	
	The SM input parameters are taken as follows~\cite{ParticleDataGroup:2024cfk}:
	\begin{equation}
		\begin{aligned}
			m_h &= 125.20~\mathrm{GeV}, &\quad m_Z &= 91.1880~\mathrm{GeV}, &\quad &\alpha_s(m_Z) = 0.1180, \\
			m_t &= 172.57~\mathrm{GeV}, &\quad m_W &= 80.3692~\mathrm{GeV}, &\quad &\alpha = 1/137.035999.
		\end{aligned}
		\label{eq:inputs}
	\end{equation}
	The strong coupling constant $\alpha_s(\mu)$ is evolved from $\mu = m_Z$ to the relevant energy scale $\mu = \sqrt{s}$ by means of the package \texttt{RunDec}~\cite{Chetyrkin:2000yt,Herren:2017osy}.
	
	The parameter space is scanned over the following ranges:
	\begin{equation}
		\begin{aligned}
			m_H = m_A = m_{H^{\pm}} \in [126, 1000]~\mathrm{GeV},\quad M^2 \in [0, 1 \times 10^6]~\mathrm{GeV}^2,\quad \tan\beta \in [1.5, 50].
		\end{aligned}
	\end{equation}
	We first consider the alignment limit, i.e. $s_{\beta-\alpha}=1$. A sample of 5000 parameter points satisfying all theoretical and experimental constraints detailed in Sec.~\ref{sec:model} is retained.
	
	\begin{figure*}
		\centering
		\includegraphics[scale=0.28]{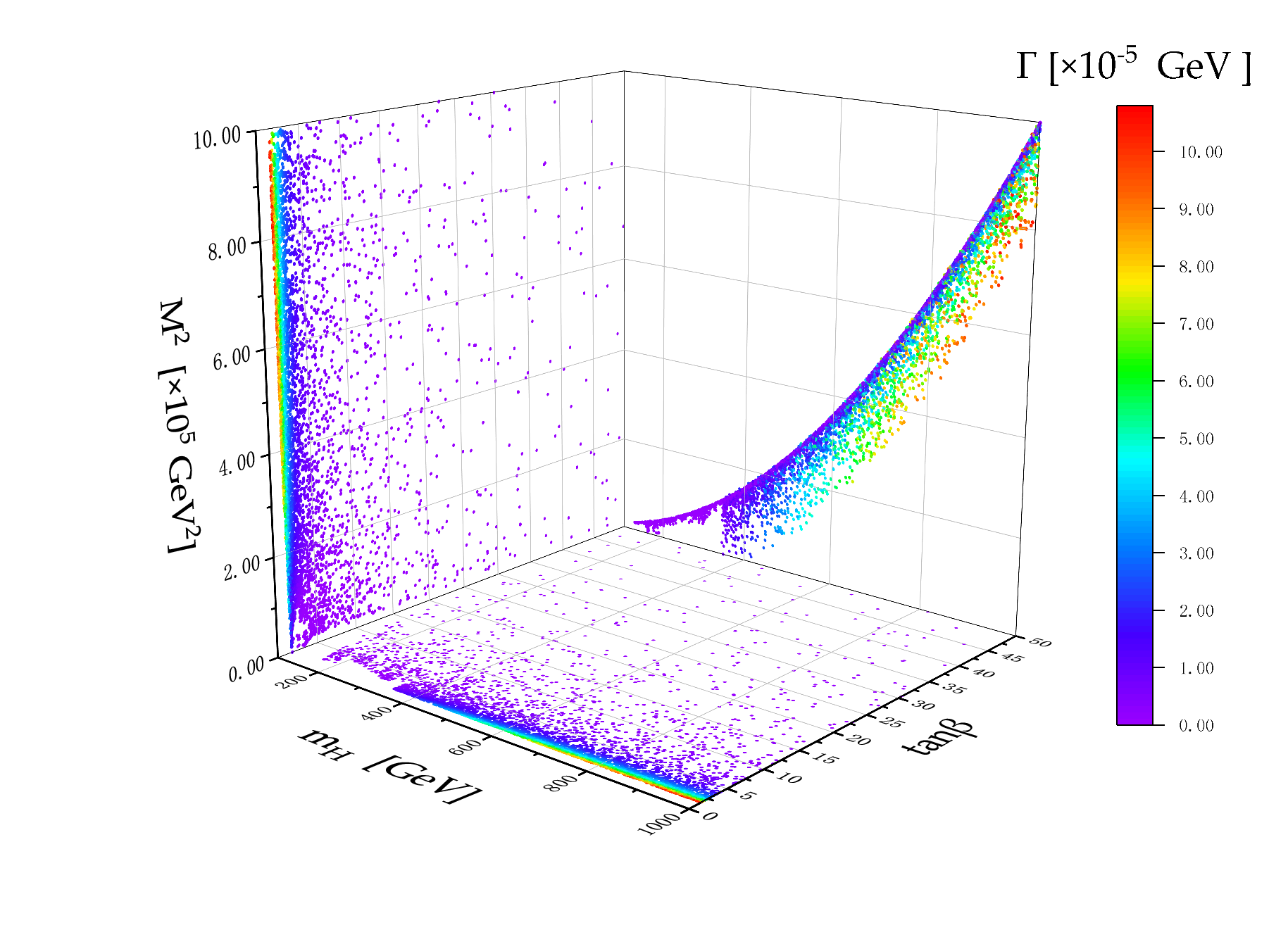}
		\includegraphics[scale=0.28]{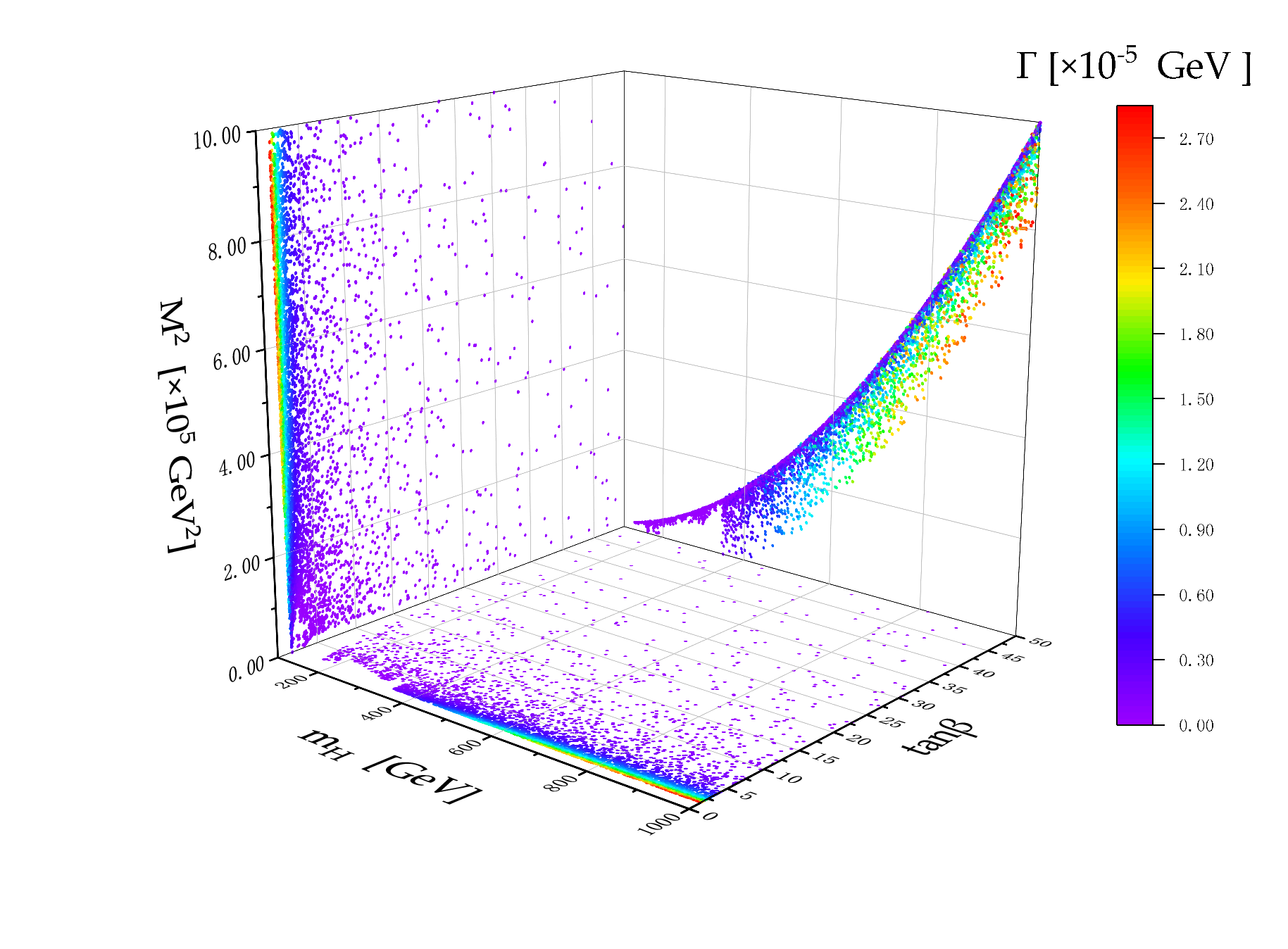}
		\caption{Decay widths for $H \to \gamma\gamma$ (left panel) and $H \to Z\gamma$ (right panel) at LO projected onto different parameter planes, for the 5000 viable parameter points in the alignment limit ($s_{\beta-\alpha}=1$). \label{fig:lo}}
	\end{figure*}
	
	In Figure~\ref{fig:lo}, the LO decay widths for the two processes are projected onto various parameter planes. The results for $H \to \gamma\gamma$ and $H \to Z\gamma$ are presented in the left and right panels, respectively. The viable points cluster predominantly at small values of $\tan\beta$, with $M^2$ approximately proportional to $m_H^2$, a feature driven primarily by theoretical constraints. The irregular structure in the region $m_H \sim 200$--$400~\mathrm{GeV}$ originates from experimental constraints on $H^+ \to q q^{\prime}$~\cite{CMS:2020imj,ATLAS:2021upq} and $\phi \to \gamma\gamma / \tau\tau$~\cite{ATLAS:2017ayi,ATLAS:2020zms}, consistent with the behaviour reported in Ref.~\cite{Arbey:2017gmh}. The decay widths of both processes display analogous patterns across the projected planes. With increasing neutral Higgs mass, the decay widths tend to grow owing to the enlarged phase space. Overall, the decay width for $H \to \gamma\gamma$ systematically exceeds that for $H \to Z\gamma$. For $m_H \lesssim 400~\mathrm{GeV}$ and $M^2 < 8\times10^4~\mathrm{GeV}^2$, the decay widths for both processes become highly suppressed, falling below approximately $1\times10^{-5}$ and $3\times10^{-6}$ for $H \to \gamma\gamma$ and $H \to Z\gamma$, respectively.

	As also seen in Figure~\ref{fig:lo}, the decay widths diminish rapidly with increasing $\tan\beta$. This decrease is monotonic and admits a simple explanation. In the alignment limit ($c_{\beta-\alpha}=0$), the $HWW$ coupling vanishes, implying $A_W^{(\mathrm{1\text{-}loop})}=0$. The LO decay width therefore receives contributions only from the top-quark loop, the charged-Higgs loop, and their interference,
	\begin{equation}
		\Gamma_{\mathrm{LO}}
		=
		\Gamma_{\mathrm{top}}
		+
		\Gamma_{H^\pm}
		+
		\Gamma_{t\text{-}H^\pm}.
		\label{eq:LOdecomp}
	\end{equation}
	
\begin{figure}[htbp]
		\centering
		\includegraphics[width=\textwidth]{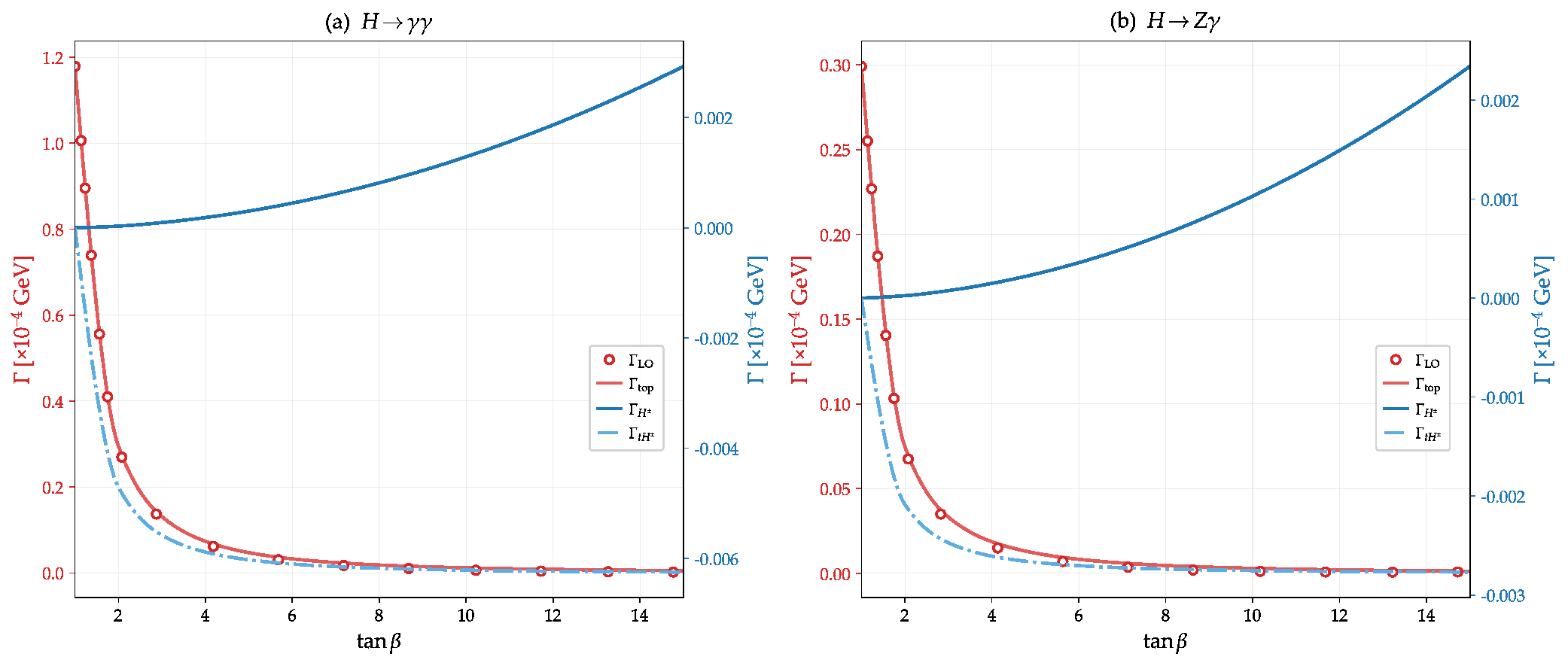}
		\caption{Decomposition of the LO decay width into top-quark ($\Gamma_{\mathrm{top}}$, red solid line), charged-Higgs ($\Gamma_{H^\pm}$, blue solid line), and interference ($\Gamma_{t\text{-}H^\pm}$, blue dashed line) contributions in the alignment limit ($s_{\beta-\alpha}=1$), with $m_H=489$~GeV and $M^2=226674$~GeV$^2$. Red curves are read on the left axis; blue curves on the right axis. The red circles show the total LO width $\Gamma_{\mathrm{LO}}$. Panel~(a): $H\to\gamma\gamma$; panel~(b): $H\to Z\gamma$.}
		\label{fig:decomposition}
	\end{figure}

	From Eqs.~\eqref{eq:mathcalA} and \eqref{eq:I1I2}, the scalar loop function $\mathcal{A}_0$ is of the same order as the fermionic one, $\mathcal{A}_{1/2}$. In the alignment limit, however, Eq.~\eqref{eq:ampVgamma} shows that the charged-Higgs contribution carries a further suppression from the effective trilinear coupling
	\begin{equation}
		\lambda_{\mathrm{eff}}
		=
		\frac{(m_H^2-M^2)\cot 2\beta}{m_{H^\pm}^2},
		\label{eq:lambdaeff}
	\end{equation}
	which stays small over most of the viable parameter space. As a result, unless $\tan\beta$ becomes very large, both the pure charged-Higgs term and the interference are numerically negligible against the top-quark loop. 
	The total decay width is therefore well approximated by
	\begin{equation}
		\Gamma_{\mathrm{LO}}
		\simeq \Gamma_{\mathrm{top}}
		\propto
		\cot^2\beta,
		\label{eq:GammaApprox}
	\end{equation}
	which naturally explains the observed monotonic decrease with increasing $\tan\beta$. 
	Figure~\ref{fig:decomposition} illustrates how each contribution scales with $\tan\beta$: $\Gamma_{\mathrm{top}}$ exceeds both $\Gamma_{H^\pm}$ and $\Gamma_{t\text{-}H^\pm}$ by several orders of magnitude over nearly the whole $\tan\beta$ range.
	Because the viable THDM parameter space clusters at moderate $\tan\beta$, this $\cot^2\beta$ scaling controls the decay widths throughout the phenomenologically relevant region.

	\begin{figure*}
		\centering
		\includegraphics[scale=0.28]{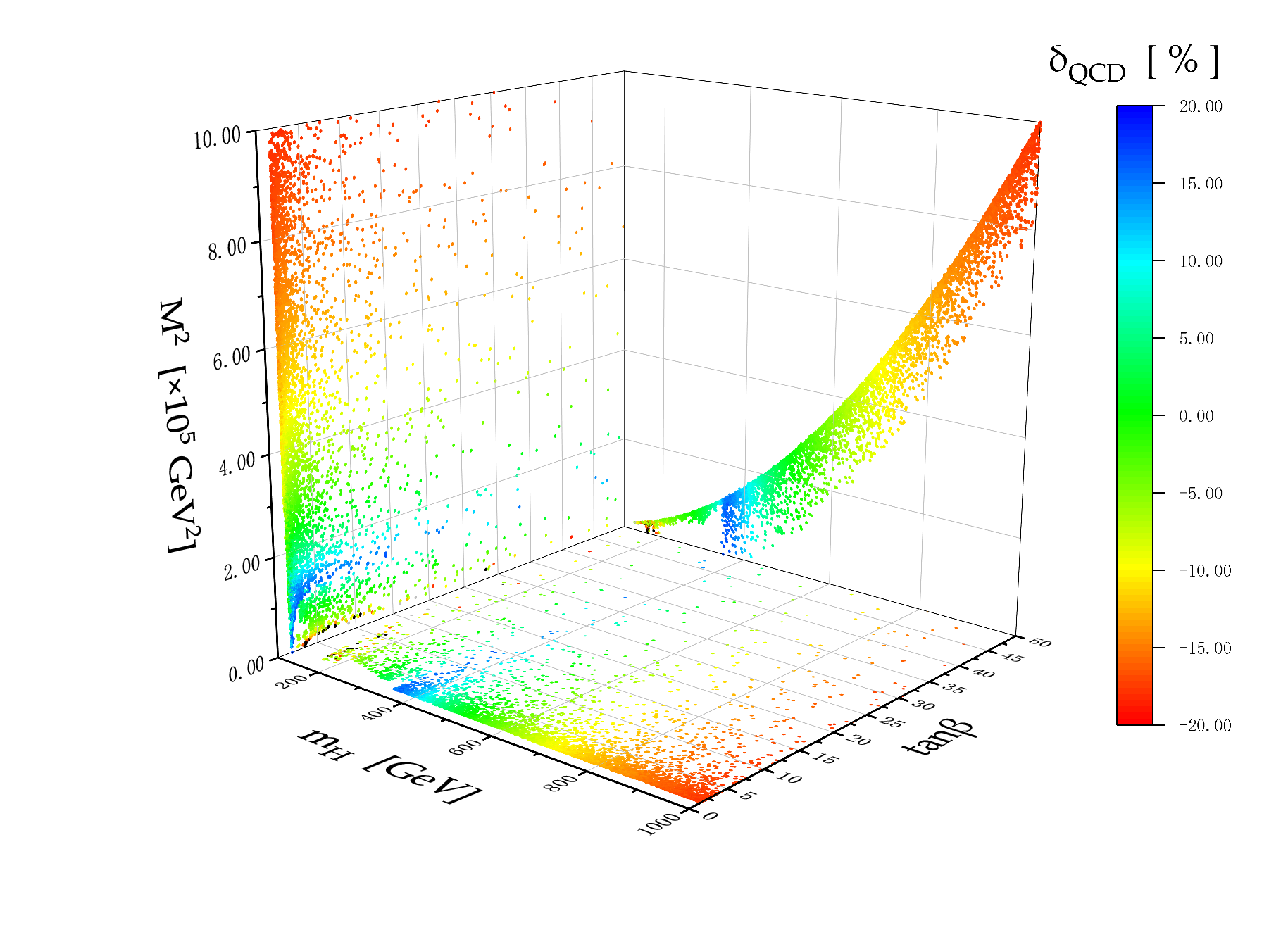}
		\includegraphics[scale=0.28]{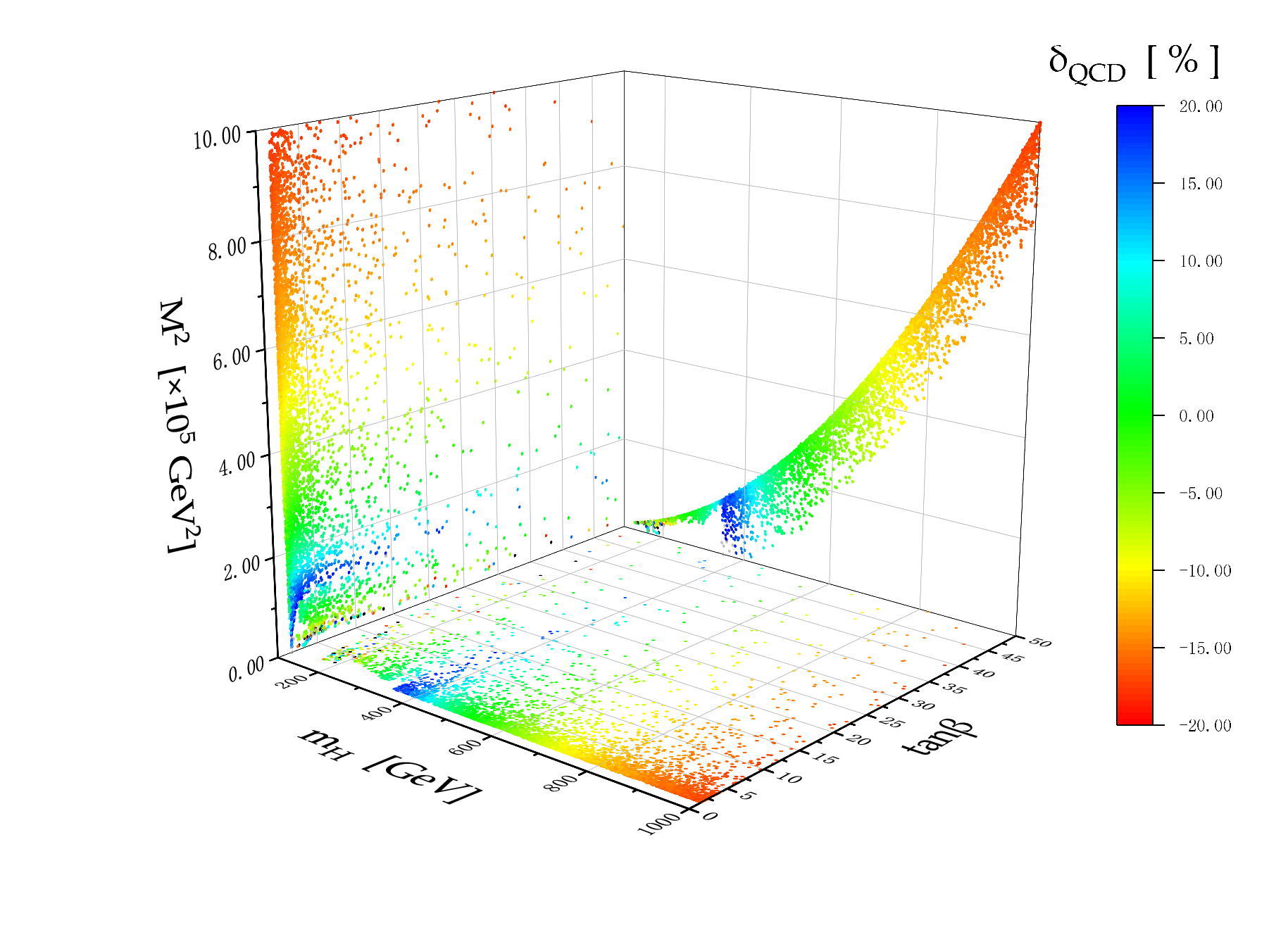}
		\caption{Relative corrections $\delta_{\mathrm{QCD}}$ for $H \to \gamma\gamma$ (left panel) and $H \to Z\gamma$ (right panel) at NLO projected onto different parameter planes. \label{fig:re}}
	\end{figure*}
	
	\begin{figure*}
		\centering
		\includegraphics[scale=0.31]{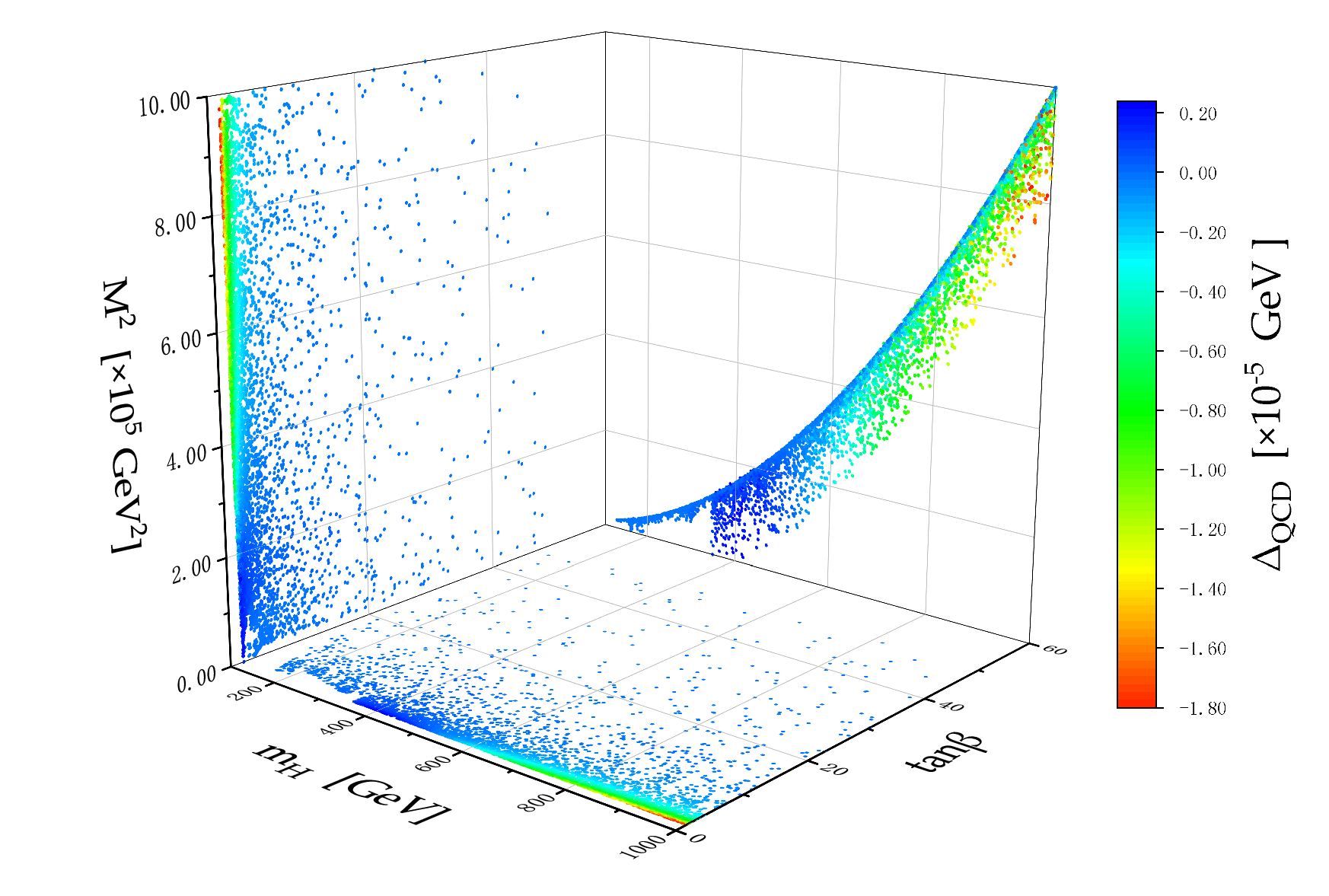}
		\includegraphics[scale=0.31]{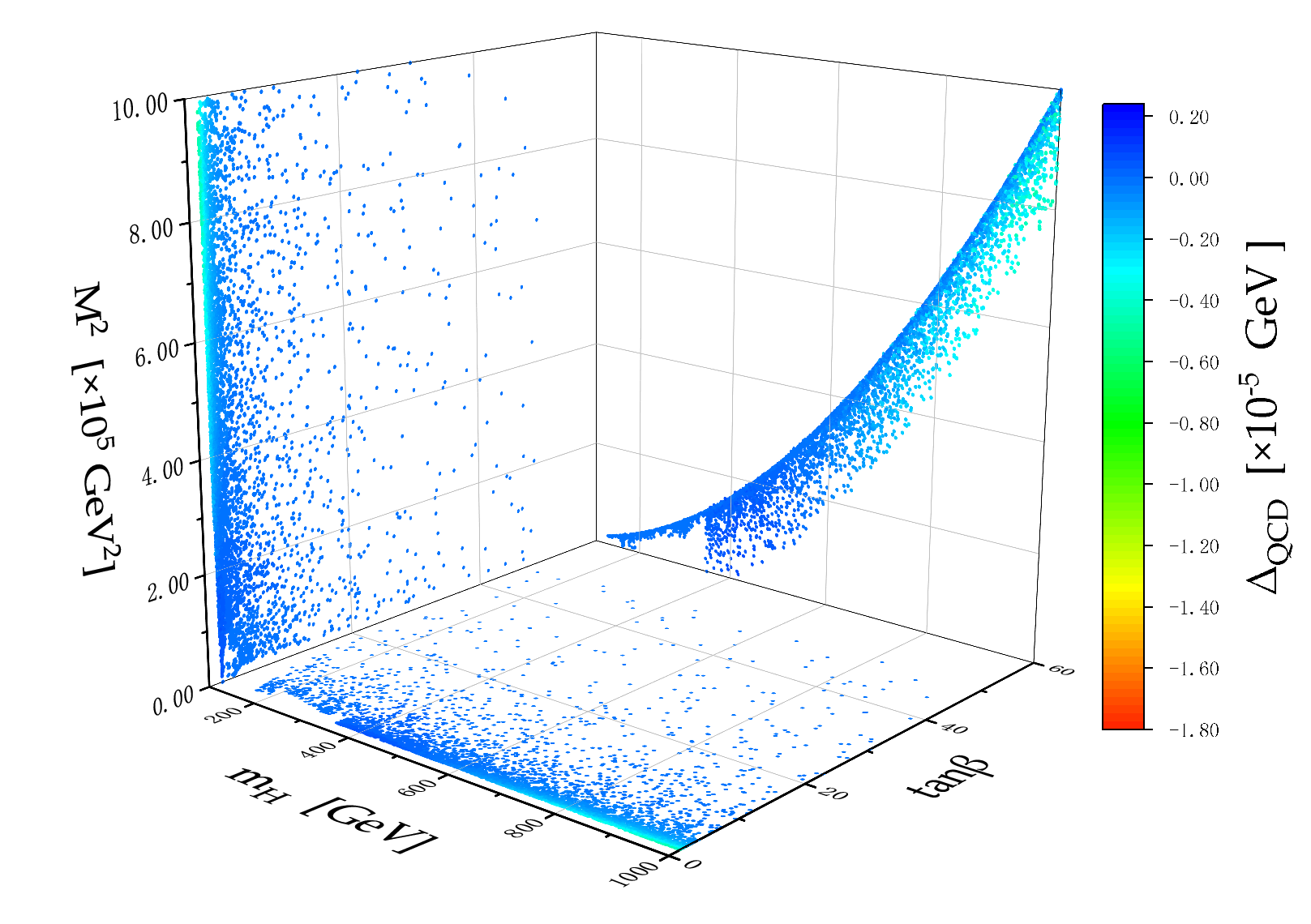}
		\caption{Absolute corrections $\Delta_{\mathrm{QCD}}$ for $H \to \gamma\gamma$ (left panel) and $H \to Z\gamma$ (right panel) at NLO projected onto different parameter planes, for the same parameter sample as in Figure~\ref{fig:lo}. \label{fig:nloabs}}
	\end{figure*}
	
	The relative corrections for the two processes are presented in Figure~\ref{fig:re}. Results for $H \to \gamma\gamma$ appear in the left panels and those for $H \to Z\gamma$ in the right panels. The relative corrections exhibit no appreciable dependence on $\tan\beta$ but depend primarily on $m_H$. On both the $m_H$--$\tan\beta$ and $m_H$--$M^2$ planes, distinct colour-coded bands corresponding to different values of $m_H$ are clearly visible. As $m_H$ increases, the relative correction first rises and subsequently falls, attaining its maximum in the vicinity of the $t\bar{t}$ production threshold. Beyond this threshold, the relative correction decreases monotonically and eventually changes sign from positive to negative. At $m_H \sim 1000~\mathrm{GeV}$, the relative correction amounts to approximately $-20\%$. At very small values of $m_H$ or $M^2$, a few isolated outliers emerge where the relative corrections become appreciably large. This behaviour is attributable to the strongly suppressed LO decay width in these regions, whereas the NLO contribution retains a moderate magnitude, as shown in Figure~\ref{fig:lo} and Figure~\ref{fig:nloabs}.
	
	Figure~\ref{fig:nloabs} displays the NLO contributions $\Delta_{\text{QCD}}$. The parametric dependence of the NLO contributions mirrors that observed at LO. The NLO corrections arise from the top-quark loop, whose Yukawa coupling scales as $\cot\beta$. The interference between the two-loop and one-loop amplitudes consequently reproduces the LO pattern.
	
	To delineate the individual parameter dependences more transparently, one-dimensional projections are presented below. Figure~\ref{fig:others} displays the decay widths and corresponding NLO QCD relative corrections as functions of $\tan\beta$ for $H \to \gamma\gamma$ (left panel) and $H \to Z\gamma$ (right panel) with $M^2 = m_H^2$. In each panel, three curves are shown corresponding to $s_{\beta-\alpha} = 0.98,\, 0.99,$ and $1$. Results are presented for $m_H = 350,\, 550,$ and $850~\mathrm{GeV}$, thereby facilitating a direct comparison of the mass dependence across different $m_H$ values.
	
	\begin{figure*}
		\centering
		\includegraphics[scale=0.30]{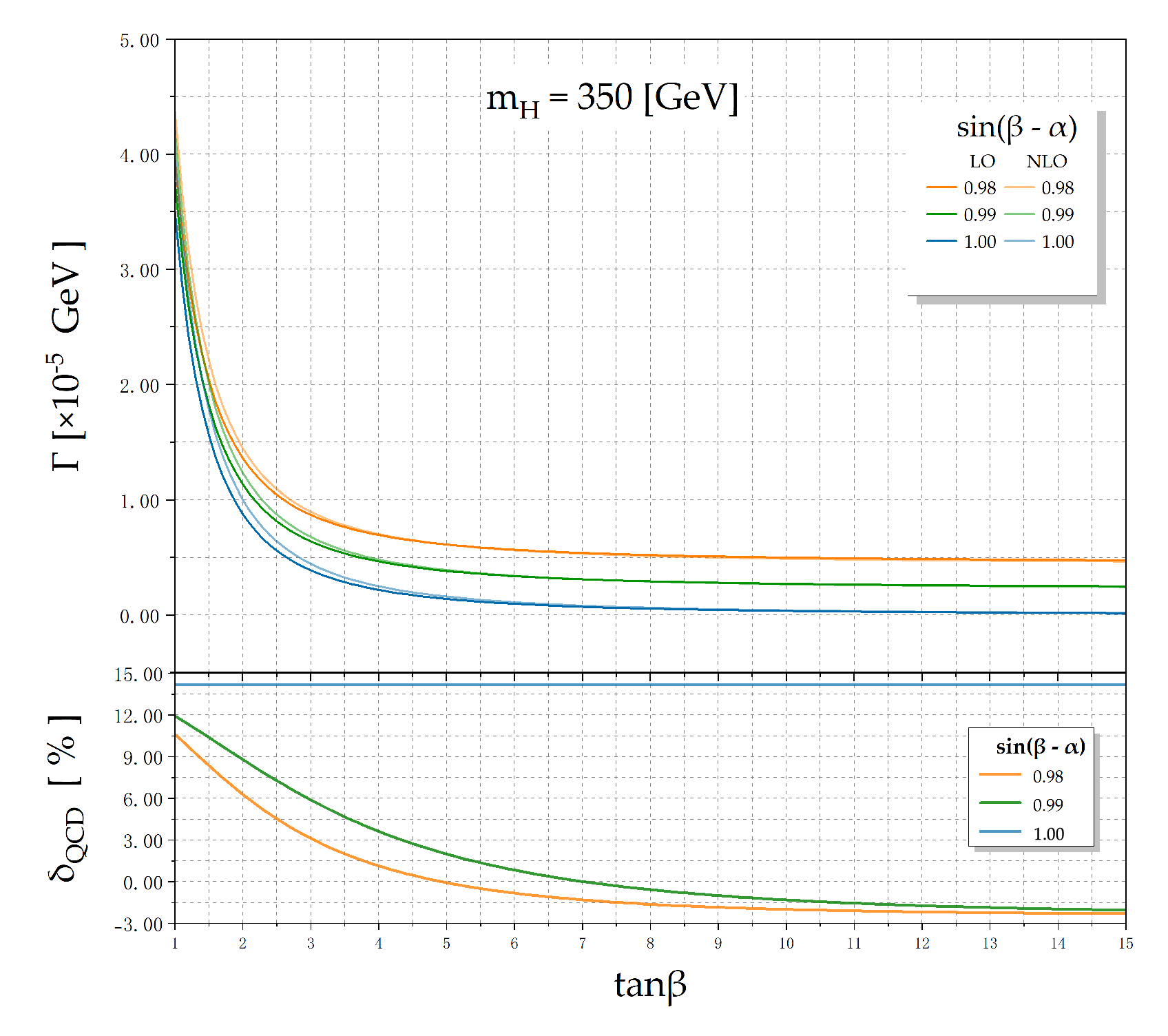}
		\includegraphics[scale=0.30]{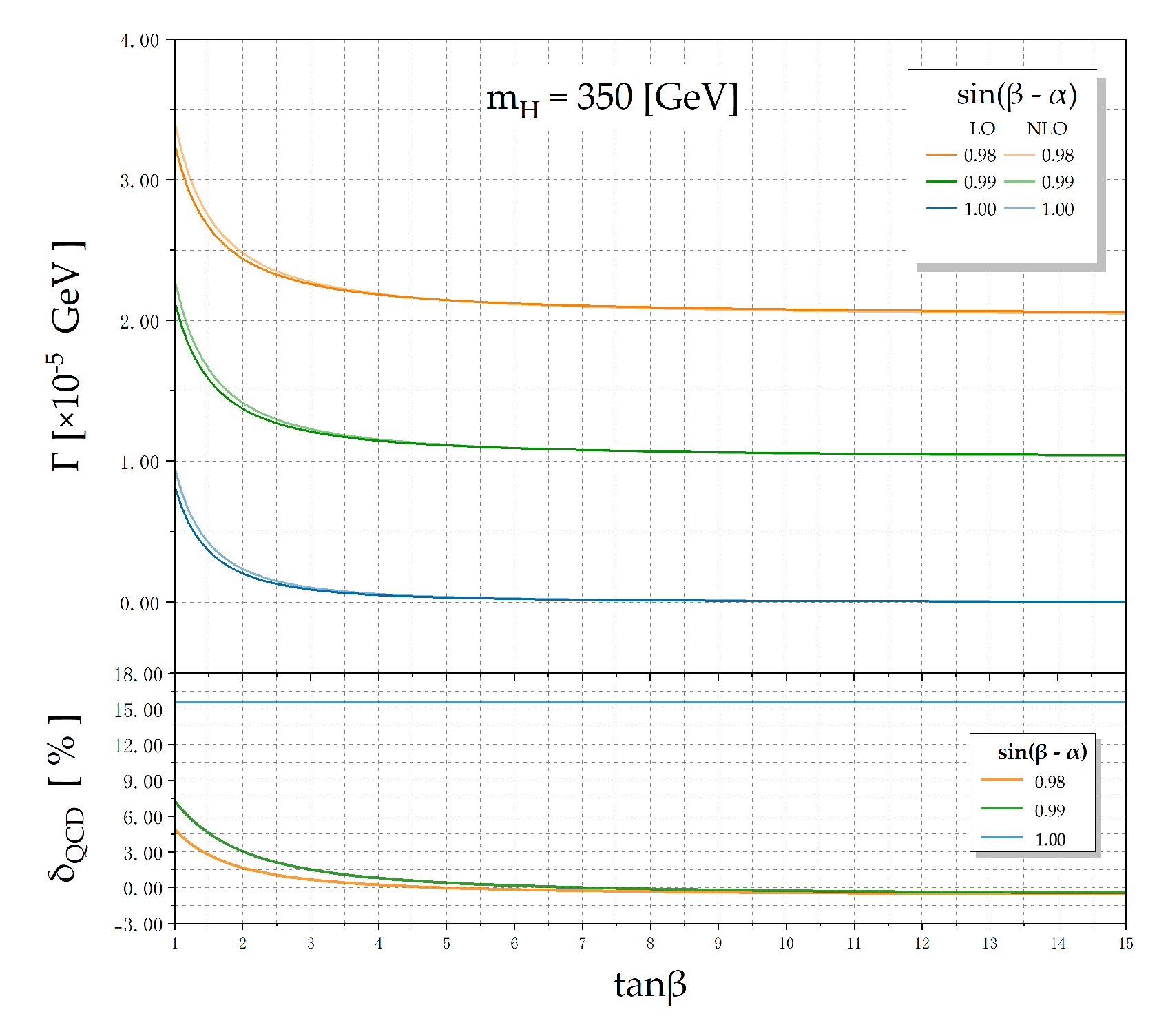}
		\includegraphics[scale=0.30]{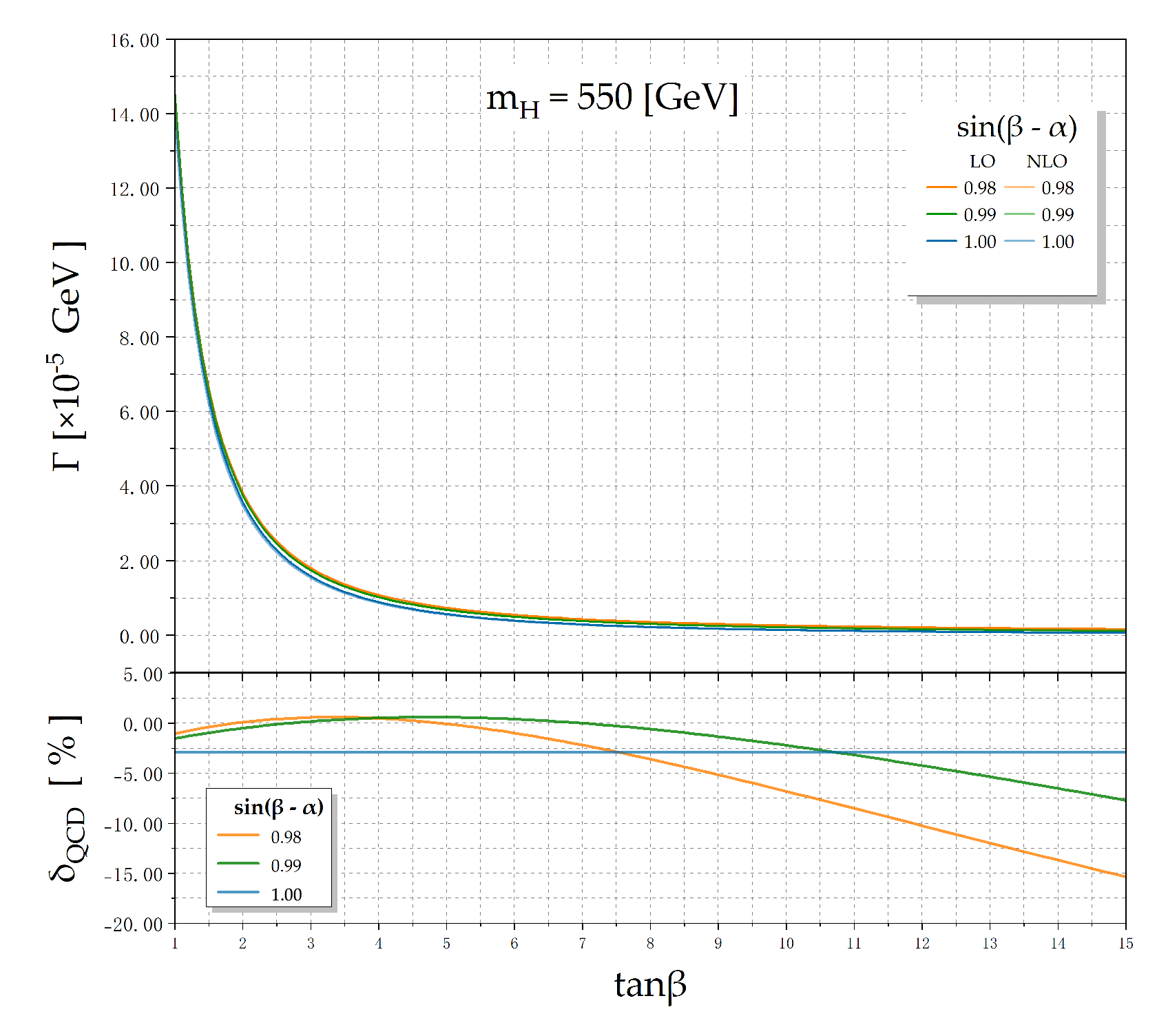}
		\includegraphics[scale=0.30]{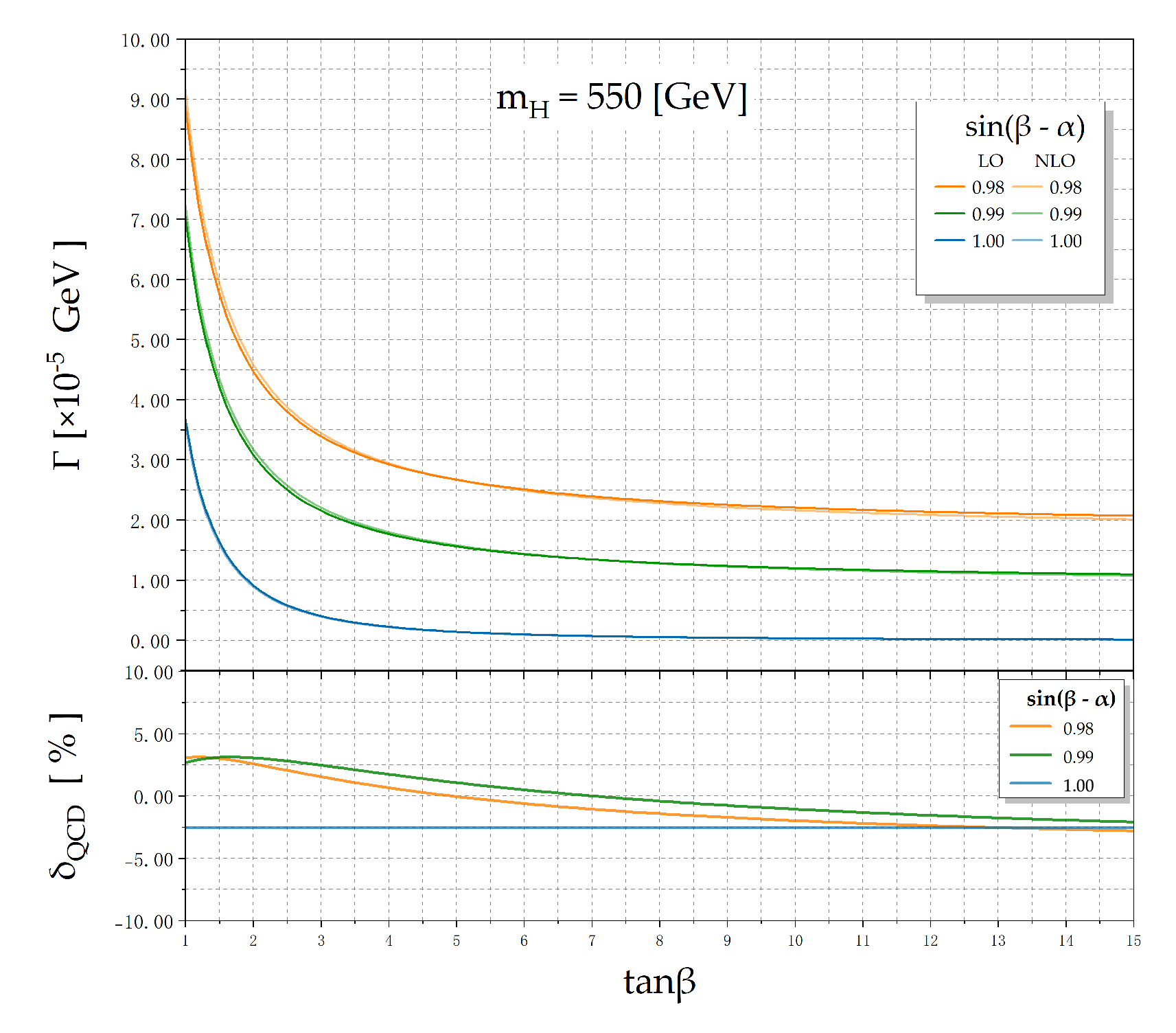}
		\includegraphics[scale=0.30]{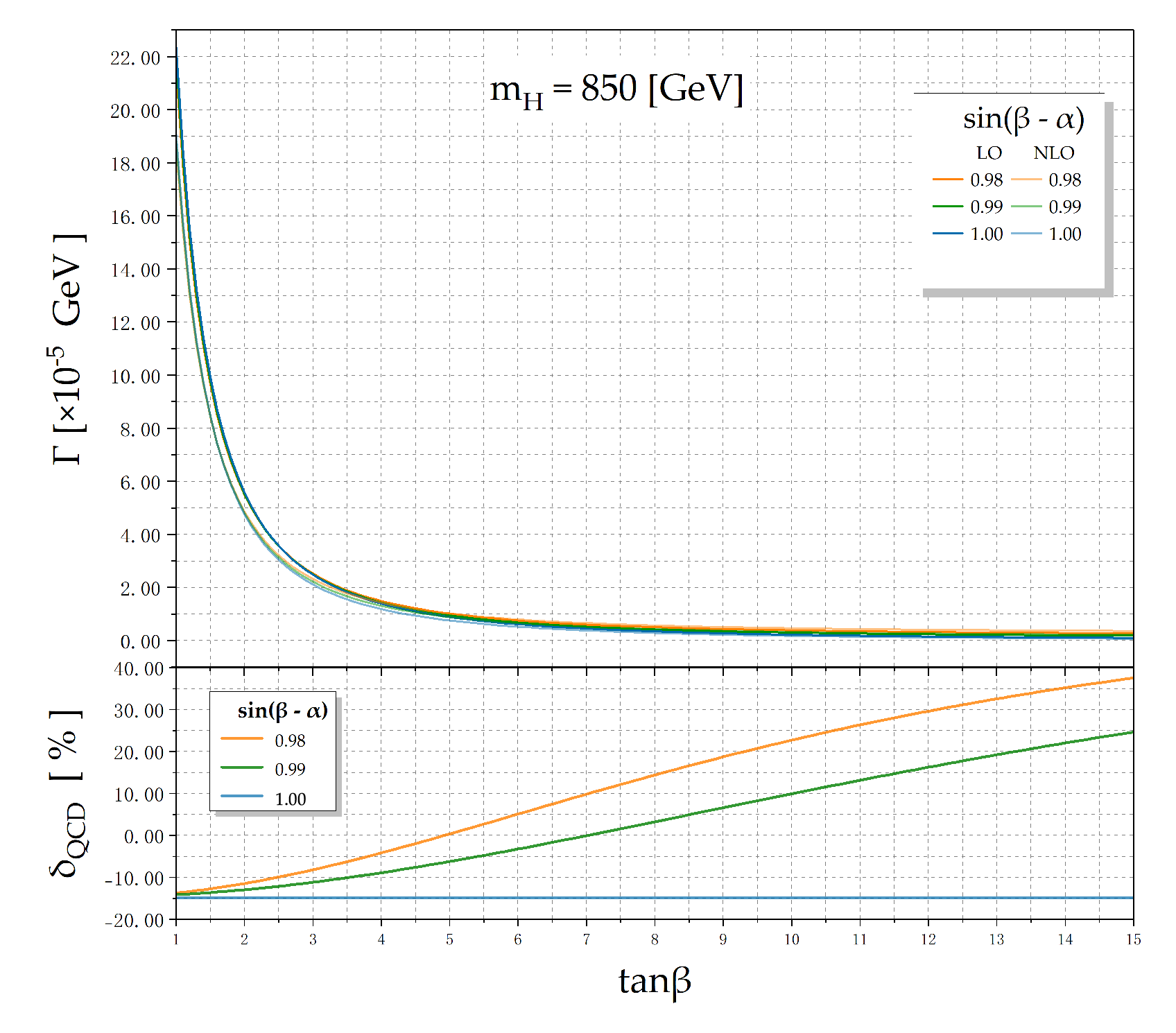}
		\includegraphics[scale=0.30]{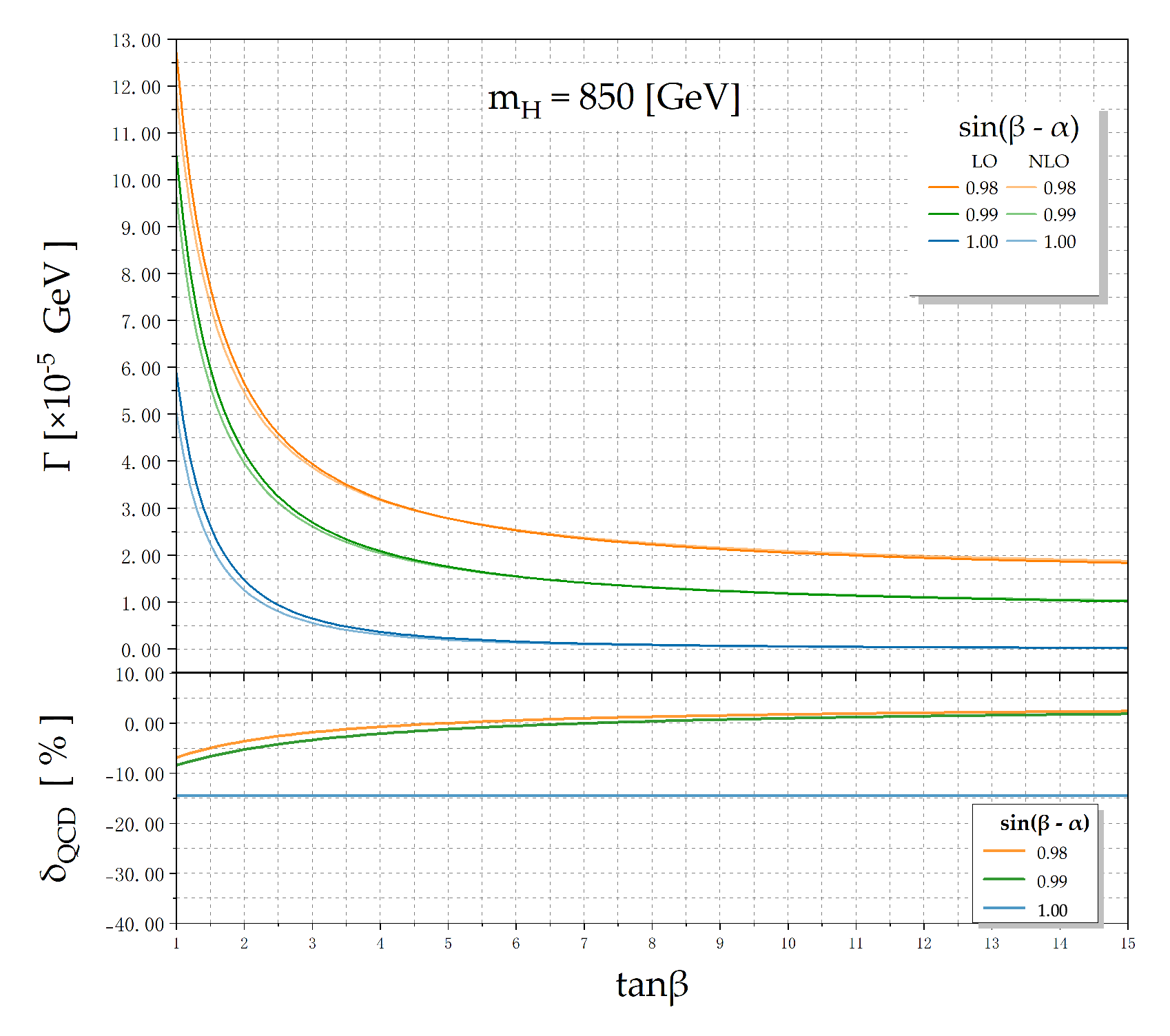}
		\caption{Decay widths and relative corrections for $H \to \gamma\gamma$ (left) and $H \to Z\gamma$ (right) as functions of $\tan\beta$ for different $m_H$ and $s_{\beta-\alpha}$ values, with $M^2=m_H^2$. From top to bottom: $m_H=350$, $550$, and $850~\mathrm{GeV}$; in each panel, the upper plot shows the decay widths at LO and NLO for $s_{\beta-\alpha}=0.98$, $0.99$, and $1$, and the lower plot the corresponding relative correction $\delta_{\mathrm{QCD}}$. \label{fig:others}}
	\end{figure*}
	
	In both channels, the decay widths diminish monotonically with increasing $\tan\beta$. In the $H \to \gamma\gamma$ channel, the three $s_{\beta-\alpha}$ curves exhibit the greatest separation at $m_H = 350~\mathrm{GeV}$. As $m_H$ increases, the inter-curve differences progressively diminish, and by $m_H = 550~\mathrm{GeV}$ the decay width curves nearly converge. In contrast, for the $H \to Z\gamma$ channel, the three curves remain discernibly separated across the full $m_H$ range, albeit with diminishing spacing as $m_H$ increases. In general, the decay width is enhanced for smaller $s_{\beta-\alpha}$, as the $HWW$ coupling is proportional to $c_{\beta-\alpha}$ and the magnitude of $\kappa_{f,H}$ grows with decreasing $s_{\beta-\alpha}$, as illustrated in Table~\ref{tab:vertices}. For $m_H = 350~\mathrm{GeV}$, the corrections with $s_{\beta-\alpha}\neq 1$ fall monotonically with increasing $\tan\beta$, crossing zero at $\tan\beta\approx 5$ ($s_{\beta-\alpha}=0.98$) and $\tan\beta\approx 7$ ($s_{\beta-\alpha}=0.99$); the $s_{\beta-\alpha}=1$ curve instead stays flat, since the $\cot^2\!\beta$ scaling cancels exactly in the ratio $\delta_{\mathrm{QCD}}$. At $m_H = 550~\mathrm{GeV}$, the $s_{\beta-\alpha}\neq 1$ curves turn mildly non-monotonic. In the $H\to\gamma\gamma$ channel they begin slightly negative at small $\tan\beta$, rise to a shallow positive maximum of about $0.6\%$ near $\tan\beta\approx 3$--$5$, and then decrease to progressively more negative values. In the $H\to Z\gamma$ channel they start positive (around $3\%$), peak near $\tan\beta\approx 1$--$2$, and subsequently decrease, eventually turning negative at large $\tan\beta$. By contrast, at $m_H = 850~\mathrm{GeV}$ the corrections with $s_{\beta-\alpha}\neq 1$ rise monotonically with $\tan\beta$ in both channels, reaching about $+38\%$ for $s_{\beta-\alpha}=0.98$ in the $H\to\gamma\gamma$ channel at large $\tan\beta$. In the $H\to Z\gamma$ channel these curves likewise cross zero into positive territory, attaining roughly $+2\%$ for $s_{\beta-\alpha}=0.98$.

	\subsection{QCD effects in the decoupling limit\label{sec:decoupling}}
	\begin{figure*}
		\centering
		\includegraphics[scale=0.31]{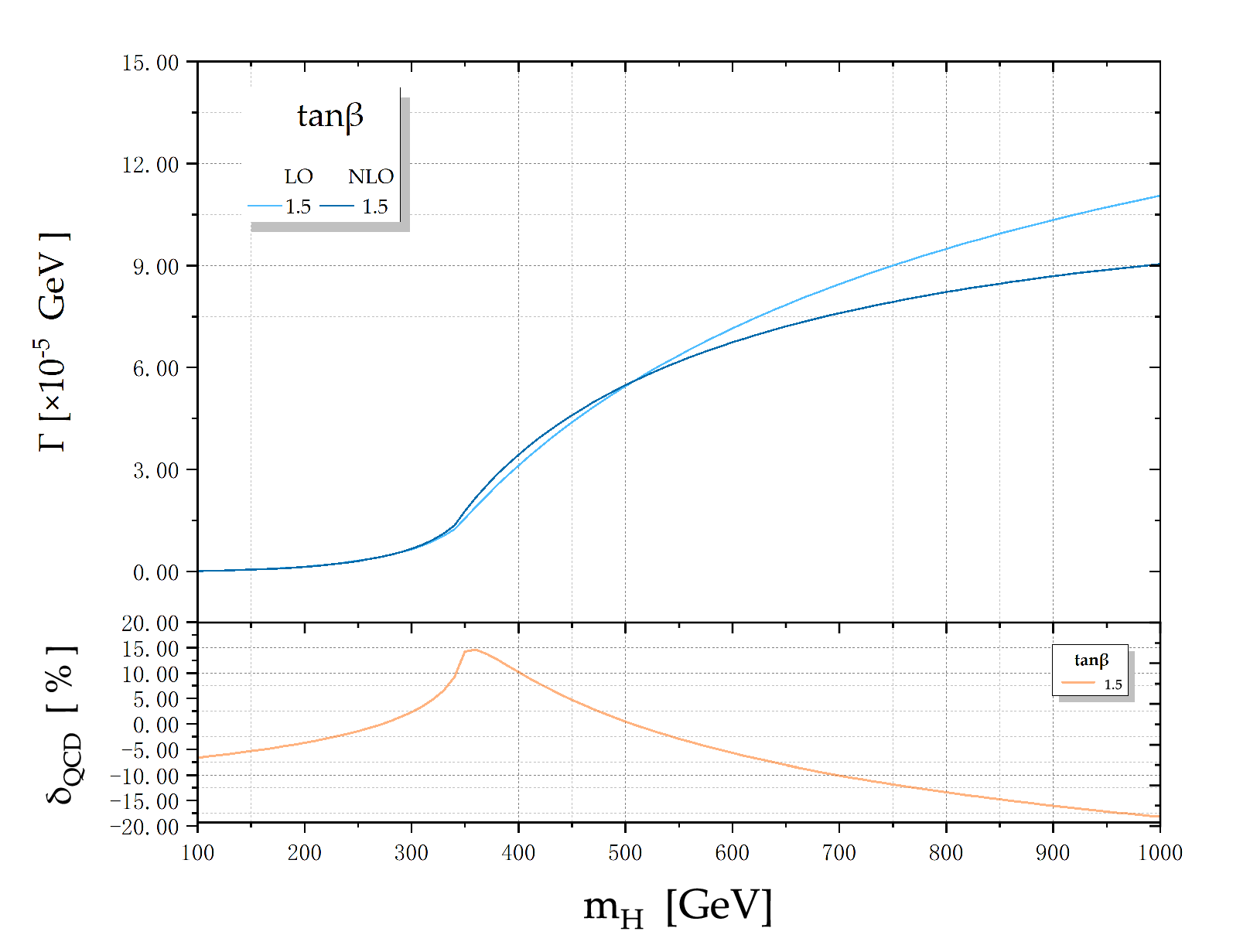}
		\includegraphics[scale=0.31]{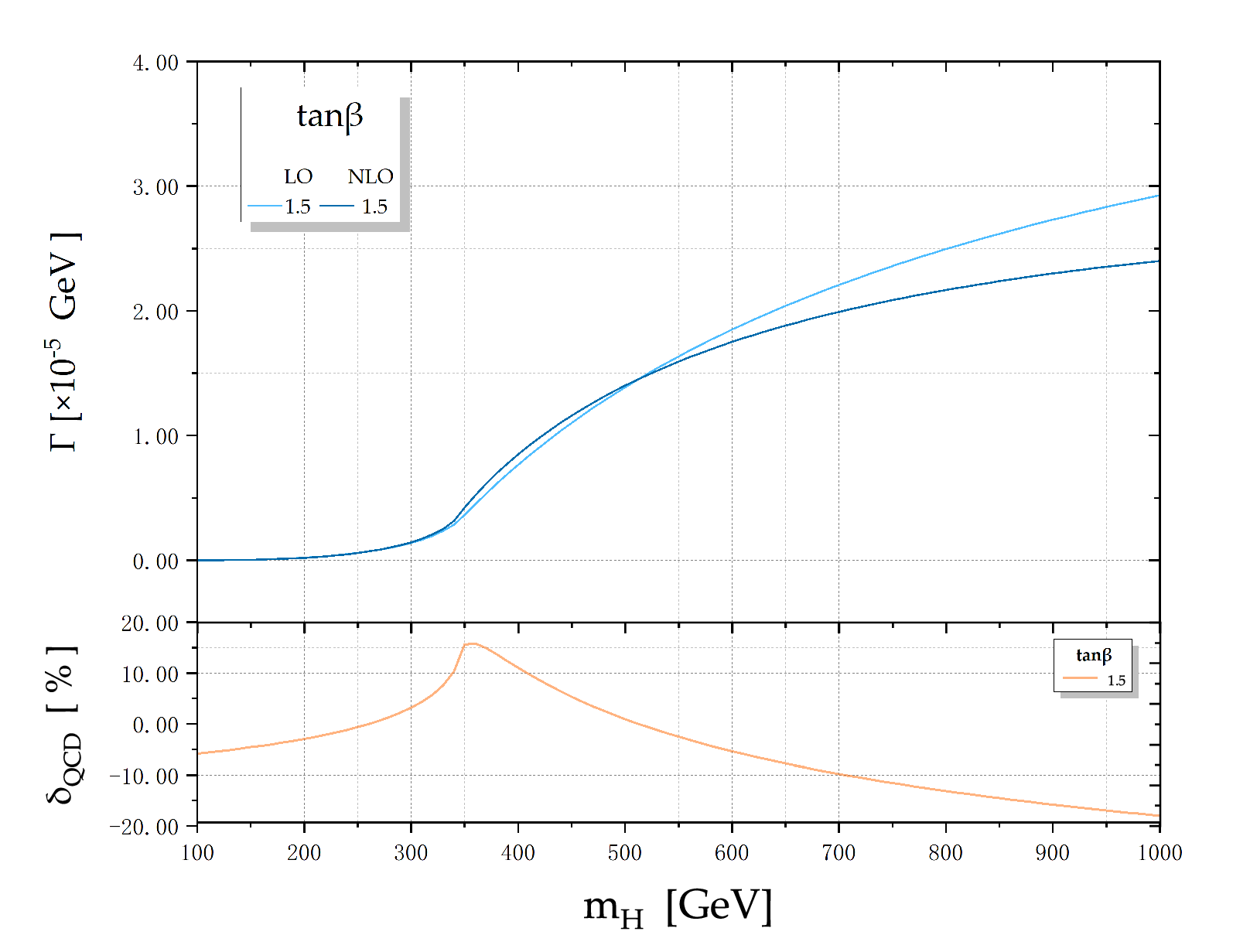}
		\caption{Decay widths and relative corrections for $H \to \gamma\gamma$ (left) and $H \to Z\gamma$ (right) as functions of $m_H$ for $\tan\beta=1.5$, in the decoupling limit ($s_{\beta-\alpha}=1$ and $M^2=m_H^2$). In each panel, the upper plot shows the decay widths at LO and NLO, and the lower plot the relative correction $\delta_{\mathrm{QCD}}$. \label{fig:nloabs-mh}}
	\end{figure*}
	The QCD corrections to the amplitude originate solely from top-quark loops. A top-only analysis therefore already captures the essential features, in particular the neutral-Higgs mass dependence seen in Figure~\ref{fig:lo}. We accordingly define the decoupling limit as the regime in which only the top-quark loop contributes, which corresponds to $s_{\beta-\alpha}=1$ together with $m_{H}^2=M^2$.
	In Figure~\ref{fig:nloabs-mh}, we depict the decay widths and QCD relative corrections for the $H \to \gamma\gamma$ and $H \to Z\gamma$ channels in this case, choosing $\tan\beta = 1.5$ as a function of the heavy Higgs mass $m_H$. Within the $m_H$ range considered, the decay widths and relative corrections for both channels exhibit closely analogous behaviour. The decay widths grow with $m_H$. Quantitatively, the $H \to \gamma\gamma$ decay width exceeds that of $H \to Z\gamma$ by roughly a factor of four. The QCD relative corrections range from $-20\%$ to $+20\%$, following nearly identical trends in the two channels. A pronounced peak is evident near the $t\bar{t}$ production threshold, where the corrections attain approximately $15\%$. Beyond this threshold, the QCD corrections turn negative and continue to diminish with increasing $m_H$. 
	
	The pronounced enhancement near the $t\bar t$ threshold in Fig.~\ref{fig:nloabs-mh} prompts the question of whether the fixed-order QCD prediction remains reliable there. Unlike processes dominated by an $S$-wave heavy-quark pair, the threshold behaviour of the scalar Higgs decay amplitudes is governed by a $P$-wave configuration, for which Coulomb effects are strongly suppressed~\cite{Melnikov:1994jb}. This suppression follows directly from the quantum numbers of the intermediate $t\bar t$ state.
	
	For a $CP$-even Higgs boson, the Yukawa interaction $y_t\bar t tH$ conserves $CP$, requiring the intermediate $t\bar t$ pair to have quantum numbers
	\begin{equation}
		J^{PC}=0^{++}.
	\end{equation}
	The lowest $S$-wave configuration ${}^{1}S_{0}$ instead carries
	\begin{equation}
		J^{PC}=0^{-+},
	\end{equation}
	and is therefore forbidden by $CP$ conservation. Consequently, the lowest allowed threshold state is the ${}^{3}P_{0}$ configuration, implying that the production amplitude is proportional to the relative momentum of the heavy quarks.
	
	Near threshold, multiple Coulomb-gluon exchanges are customarily resummed into the Sommerfeld factor \cite{Melnikov:1994jb,Sakharov:1948plh,Sommerfeld1939},
	\begin{equation}
		\Im [A_{\text{top}}]=\Im[A_{\text{top}}^{\text{{1-loop}}}]C_{\mathrm{Coul}}\,,\label{eq:Coul}
	\end{equation}
	where 
	\begin{equation}
		C_{\mathrm{Coul}}
		=
		\frac{C_F\frac{2\pi\alpha_s}{\beta_t}}
		{1-\exp\!\left(-C_F\frac{2\pi\alpha_s}{\beta_t}\right)},
	\end{equation}
	and $\beta_t=\sqrt{1-4m_t^2/m_H^2}$ denotes the top-quark velocity. Since the real part of the amplitude can be reconstructed from its imaginary part through a dispersion relation, the threshold behaviour is most transparently understood by first examining the imaginary part. 
	
	Taking $H\to\gamma\gamma$ as an example, whose amplitude reads
	\begin{equation}\label{eq:A4gammagamma_align}
		A_4(H\to \gamma\gamma)
		= \frac{\alpha}{2\pi v}\,N_c\,Q_t^2\,
		\,\kappa_{u,H}\,\left(
		\mathcal{A}_{1/2}\,
		+\frac{\alpha_s}{\pi} C_F \mathcal{A}_{1/2}^{(\text{2-loop})}
		\right),
	\end{equation}
	where $\mathcal{A}_{1/2}^{\text{(2-loop)}}$ is taken from Refs.~\cite{Harlander:2005rq,Fleischer:2004vb} and the full expression is given in Eq.~\eqref{eq:2loopA} of Appendix~\ref{sec:LOexpress}.
	The one-loop form factor admits the threshold expansion
	\begin{equation}\label{eq:Ahalf_threshold}
		\mathcal{A}_{1/2}
		=
		2
		+
		\left(
		\frac{\pi^2}{2}-2
		\right)\beta_t^2
		+
		2i\pi\beta_t^3
		+
		\mathcal{O}(\beta_t^4),
	\end{equation}
	which immediately gives
	\begin{equation}
		\mathrm{Im}\,
		\mathcal{A}_{1/2}
		=
		2\pi\beta_t^3
		+
		\mathcal{O}(\beta_t^5).
	\end{equation}

	As shown in Ref.~\cite{Melnikov:1994jb}, the Coulomb correction factor $C_{\text{Coul}}$ modifies the imaginary part of a $P$-wave amplitude only through terms that vanish at threshold,
	whereas the corresponding correction to an $S$-wave process approaches a finite constant owing to the familiar Sommerfeld enhancement. 
	Eq.~\eqref{eq:Coul} then implies a $\beta_t^2$ suppression of the Coulomb-corrected imaginary part.
	Through the dispersion relation, the real part likewise receives only finite corrections and varies smoothly across the threshold, in contrast to the logarithmic singularity characteristic of an $S$-wave amplitude. Consequently, the Coulomb effect produces only a negligible modification to the threshold behaviour of the present processes. These features already suggest that the fixed-order prediction stays well behaved across the threshold.
	
	We now expand the result explicitly near threshold at $\mathcal{O}(\alpha_s)$. For the $H\to\gamma\gamma$ amplitude, the threshold expansion of the two-loop form factor is
	\begin{equation}
		\mathcal{A}_{1/2}^{(\mathrm{2\mbox{-}loop})}
		=
		\widetilde F_0
		+
		\left(
		\widetilde c_2
		-
		2\pi^2\ln\beta_t
		+
		i\pi^3[2\Theta(\beta_t^2)-1]
		\right)
		\beta_t^2
		+
		\mathcal{O}(\beta_t^3),
	\end{equation}
	where the threshold constant is
	\begin{equation}
		\widetilde{F}_0
		= \tfrac{1}{4}\bigl(20 - 3\pi^2 + 4\pi^2\ln 4 - 28\zeta_3\bigr)\approx 2.87\,,
	\end{equation}
	$\widetilde{c}_2 = -8 + \tfrac{9}{2}\pi^2 - 7\pi^2\ln 2$ is a real constant, and $\Theta$ denotes the Heaviside step function. Compared with the LO expansion in Eq.~(\ref{eq:Ahalf_threshold}), this expansion shows that the leading contribution to the real part remains finite at threshold, while the first non-analytic correction is proportional to $\beta_t^2\ln\beta_t$. Meanwhile, the imaginary part satisfies
	\begin{equation}
		\mathrm{Im}\,
		\mathcal{A}_{1/2}^{(\mathrm{2\mbox{-}loop})}
		=
		\pi^3\beta_t^2,
	\end{equation}
	which is consistent with Eq.~\eqref{eq:Coul}. Therefore, the two-loop QCD correction preserves the same qualitative threshold structure already present at leading order: no Coulomb singularity develops, and the enhancement observed in Fig.~\ref{fig:nloabs-mh} should be interpreted as a genuine perturbative threshold effect rather than a manifestation of the breakdown of fixed-order perturbation theory. Moreover, the relative correction at the threshold is
	\begin{equation}
		\delta_{\mathrm{QCD}}(\beta_t{=}0)
		= \frac{8\alpha_s}{3\pi}\,\Re\left[\frac{\widetilde{F}_0}{F_0}\right]\,,
		\qquad F_0 = \mathcal{A}_{1/2}(\beta_t{=}0) = 2\,,
	\end{equation}
	which is approximately $1.22\alpha_s$, indicating well-behaved perturbative convergence near the threshold.
	
	Strictly speaking, in the asymptotic threshold region where $\beta_t\sim\alpha_s$, higher-order non-relativistic effects can still be systematically incorporated through threshold resummation. However, owing to the absence of Sommerfeld enhancement in the ${}^{3}P_{0}$ channel, these contributions provide only finite quantitative modifications and do not alter the analytic structure established above. We therefore expect the fixed-order calculation presented here to provide a reliable description of the overall magnitude, sign, and threshold behaviour of the QCD corrections, although the detailed line shape in an extremely narrow region around $m_H\simeq2m_t$ may receive moderate higher-order refinements.

	To assess the residual perturbative uncertainty of the NLO predictions presented above, we vary the renormalisation scale in the range $\mu\in[m_H/2,\,2m_H]$. In our calculation the strong coupling constant is renormalised in the $\overline{\mathrm{MS}}$ scheme while the quark mass is renormalised on-shell, so that the $\mu$ dependence at NLO enters only through the coupling,
	\begin{equation}
		\Gamma_{\mathrm{NLO}}(\mu)=\Gamma_{\mathrm{LO}}+\alpha_s(\mu)\,\Delta\,,
	\end{equation}
	with $\Delta=\Delta_{\mathrm{QCD}}/\alpha_s(\mu)$ and $\Gamma_{\mathrm{LO}}$ independent of $\mu$. Defining the scale-variation uncertainty as
	\begin{equation}\label{eq:epsilon_def}
		\varepsilon\equiv
		\frac{\left\vert\Gamma_{\mathrm{NLO}}(\mu/2)-\Gamma_{\mathrm{NLO}}(2\mu)\right\vert}{\Gamma_{\mathrm{NLO}}(\mu)}
		= R(\mu)\,\frac{\left\vert\delta_{\mathrm{QCD}}\right\vert}{1+\delta_{\mathrm{QCD}}}\,,
		\qquad
		R(\mu)\equiv\frac{\alpha_s(\mu/2)-\alpha_s(2\mu)}{\alpha_s(\mu)}\,,
	\end{equation}
	and using the one-loop running $\alpha_s(k\mu)\simeq\alpha_s(\mu)-\beta_0\,\alpha_s^2(\mu)\ln k$ with $\beta_0=(33-2n_f)/(6\pi)$, one obtains $R(\mu)\simeq 2\beta_0\,\alpha_s(\mu)\ln 2\simeq 1.69\,\alpha_s(\mu)$ for $n_f=5$. In the range $126~\mathrm{GeV}<\mu<1000~\mathrm{GeV}$, $\alpha_s(\mu)\simeq 0.09$--$0.11$, giving $R(\mu)\simeq 0.15$--$0.19$. Since the relative corrections satisfy $|\delta_{\mathrm{QCD}}|\lesssim 20\%$ over the viable parameter space, the resulting perturbative uncertainty is bounded by $\varepsilon\simeq 0$--$3.8\%$. The factor $1/(1+\delta_{\mathrm{QCD}})$ does enhance $\varepsilon$ for negative corrections; however, these occur only at large $m_H$, where $\alpha_s(\mu)$ (and hence $R(\mu)$) is smallest, so the bound still holds. Representative values in the decoupling limit ($s_{\beta-\alpha}=1$, $M^2=m_H^2$) are collected in Table~\ref{tab:QCDband}; the largest uncertainty, $\varepsilon=3.1\%$ at $m_H=1000~\mathrm{GeV}$, lies safely within this estimate, and the $m_H$ dependence of the tabulated values follows that of $\delta_{\mathrm{QCD}}$ shown in Figure~\ref{fig:nloabs-mh}, confirming that the NLO corrections are perturbatively stable over the full mass range considered.
	
	\begin{table}[htbp]
		\centering
		\renewcommand{\arraystretch}{1.4}
		\begin{tabular}{c cc cc}
			\toprule
			& \multicolumn{2}{c}{$H\to\gamma\gamma$}
			& \multicolumn{2}{c}{$H\to Z\gamma$} \\
			\cmidrule(lr){2-3}\cmidrule(lr){4-5}
			$m_H$ [GeV]
			& $\delta_{\mathrm{QCD}}$ [\%]
			& $\varepsilon$ [\%]
			& $\delta_{\mathrm{QCD}}$ [\%]
			& $\varepsilon$ [\%] \\
			\midrule
			300 & $+2.3$  & $0.4$ & $+3.2$  & $0.5$ \\
			350 & $+14.2$ & $2.0$ & $+15.6$ & $2.1$ \\
			500 & $+0.5$  & $0.1$ & $+0.9$  & $0.1$ \\
			700 & $-10.1$ & $1.6$ & $-9.8$  & $1.6$ \\
			1000 & $-18.2$ & $3.1$ & $-18.0$ & $3.1$ \\
			\bottomrule
		\end{tabular}
		\caption{Perturbative uncertainties induced by the scale variation $\mu\in[m_H/2,\,2m_H]$ in the alignment--decoupling limit ($s_{\beta-\alpha}=1$, $M^2=m_H^2$). $\delta_{\mathrm{QCD}}$ denotes the relative NLO correction and $\varepsilon$ the scale-variation uncertainty defined in Eq.~(\ref{eq:epsilon_def}).}
		\label{tab:QCDband}
	\end{table}
	
	\subsection{Phenomenological implications\label{sec:pheno}}
	
	\begin{figure*}
		\centering
		\includegraphics[width=\textwidth]{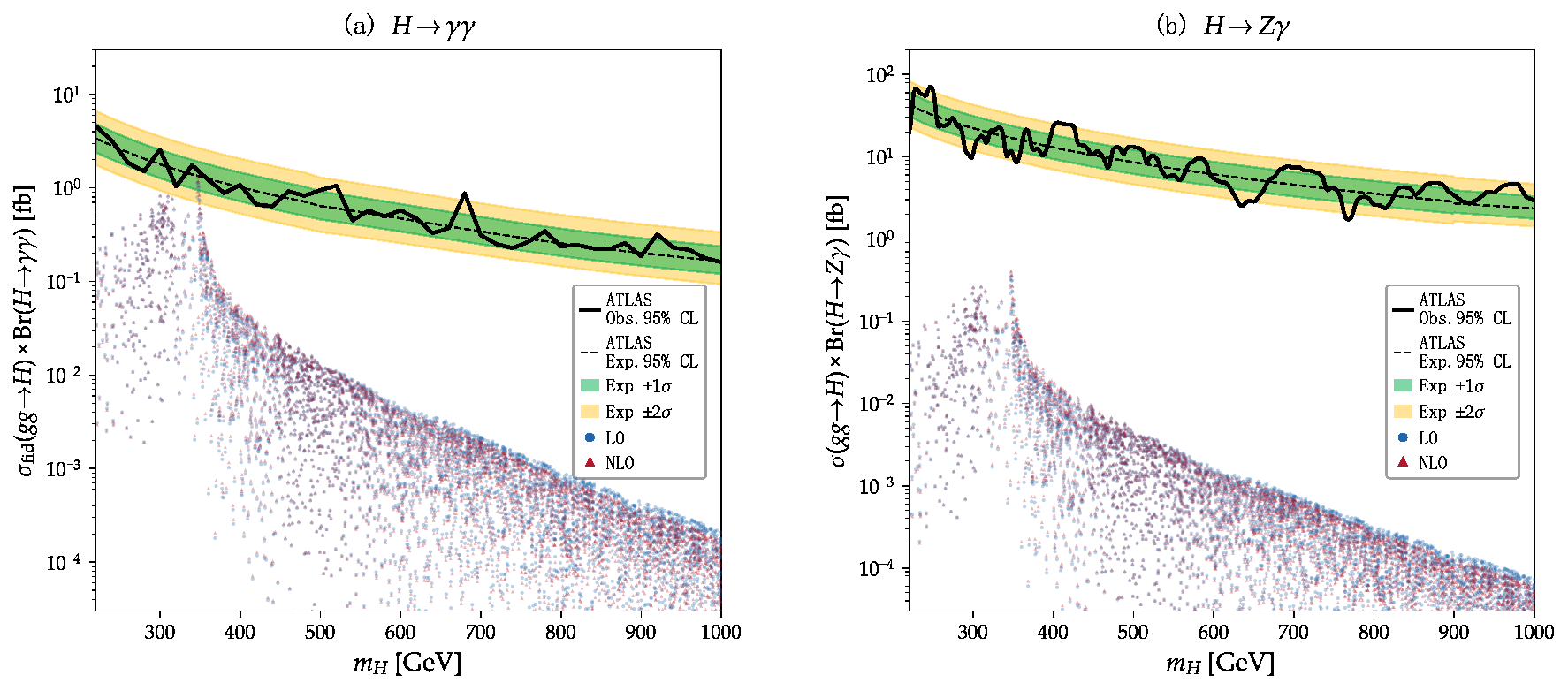}
		\caption{Predictions for $\sigma(gg\to H)\times\mathrm{Br}(H\to V\gamma)$ at LO and NLO QCD compared with the ATLAS 95\% CL upper limits for $H\to\gamma\gamma$~(a) and $H\to Z\gamma$~(b). The blue circles and red triangles denote the LO and NLO predictions for the 5000 viable parameter points, respectively; the solid and dashed black curves indicate the observed and expected limits, with the green and yellow bands representing the $\pm1\sigma$ and $\pm2\sigma$ ranges of the expected limit.
			\label{fig:pheno}}
	\end{figure*}
	
	To investigate the phenomenological implications of the NLO QCD corrections, we evaluate $\sigma(gg\to H)\times\mathrm{Br}(H\to V\gamma)$ for both the $\gamma\gamma$ and $Z\gamma$ final states and compare with existing ATLAS upper limits at $\sqrt{s}=13$~TeV. The gluon-fusion production cross section is evaluated as $\sigma(gg\to H) = \sigma_{\mathrm{SM}}^{\mathrm{NNLO+NNLL}}(m_H)\times\cot^{2}\beta$, where $\sigma_{\mathrm{SM}}^{\mathrm{NNLO+NNLL}}(m_H)$ denotes the SM Higgs boson production cross section at NNLO+NNLL accuracy as tabulated by the LHC Higgs Cross Section Working Group~\cite{deFlorian:2016spz}. In the alignment limit of the Type-I THDM, every fermion coupling to $H$ scales as $\kappa_{f,H}=-\cot\beta$; hence $\kappa_{t,H}^{2}=\cot^{2}\beta$, and the rescaling is exact at the amplitude level. The branching ratios are computed as $\mathrm{Br}(H\to V\gamma)=\Gamma(H\to V\gamma)/\Gamma_{\mathrm{total}}(H)$, where the numerator is evaluated at LO and NLO as described in Sec.~\ref{sec:processes}, and $\Gamma_{\mathrm{total}}(H)$ is the total decay width of $H$, obtained using \texttt{H-COUP}~3.0. The same $\sigma(gg\to H)$ and $\Gamma_{\mathrm{total}}(H)$ are used for both the LO and NLO predictions, so that the difference arises solely from the NLO QCD correction to $\Gamma(H\to V\gamma)$.
	
	The results are presented in Fig.~\ref{fig:pheno}. In the $H\to\gamma\gamma$ channel [panel~(a)], the predictions are compared with the ATLAS diphoton search using 139~fb$^{-1}$~\cite{ATLAS:2021uiz}. The experimental limits are reported as fiducial cross sections; accordingly, the theoretical predictions are multiplied by the acceptance 
	\begin{equation}
		\mathcal{A}_X^{ggF}(m_X)=-5.968 - 0.230\exp(-2.999\,m_{\mathrm{X}}/\mathrm{TeV}) + 6.730\exp(0.0011\,m_{\mathrm{X}}/\mathrm{TeV})
	\end{equation} 
	parametrised in the same reference. The NLO fiducial prediction reaches about 1.30~fb at $m_H\approx 347$~GeV, roughly 83\% of the observed ATLAS limit of 1.56~fb, up from 1.13~fb (73\%) at LO. The $\sim\!+15\%$ NLO correction thus narrows the gap from 0.43~fb to 0.26~fb. In the $H\to Z\gamma$ channel [panel~(b)], the predictions are compared with the ATLAS search using 140~fb$^{-1}$~\cite{ATLAS:2023wqy}. 
	The maximum $\sigma\times\mathrm{Br}$ at NLO is approximately 0.41~fb near the $t\bar{t}$ threshold ($m_H\approx 347$~GeV), roughly 30 times below the observed ATLAS limit of approximately 12~fb. The $Z\gamma$ channel thus does not yet constrain the alignment-limit parameter space. 
	With the anticipated luminosity increase at the HL-LHC, the $\mathcal{O}(15\%)$ NLO corrections may modify the exclusion boundary in the $(m_{H},\,\tan\beta)$ plane.

	\section{Conclusions \label{sec:summary}}
	
	In this work, we have investigated the decays of the heavy neutral $CP$-even Higgs boson into diphoton and $Z\gamma$ final states within the framework of the Type-I THDM. A complete numerical evaluation has been carried out retaining the full quark mass dependence. Working in the alignment limit, we have imposed the phenomenological constraints and evaluated the decay widths of both processes at NLO QCD accuracy for all viable parameter points. To elucidate the individual parameter dependences, one-dimensional projections have additionally been presented. Our findings indicate that the $H \to \gamma\gamma$ decay width typically exceeds that of $H \to Z\gamma$ by approximately a factor of four. The decay widths grow monotonically with $m_H$ owing to the enlarged phase space, while diminishing with increasing $\tan\beta$, as the dominant top-quark contribution to the width scales as $\cot^2\beta$. The LO amplitude is dominated by the top-quark loop, with the $W$-boson contribution vanishing in the alignment limit and the charged-Higgs component being suppressed across most of the viable parameter space. The QCD corrections span a range of approximately $-$20\% to $+$20\%, depending on the parameter configuration. The relative corrections exhibit a pronounced peak in the vicinity of the $t\bar{t}$ production threshold and turn negative for heavier Higgs boson masses. Away from the alignment limit, the relative corrections in the $H \to \gamma\gamma$ channel can attain values as large as $+$38\% at elevated $\tan\beta$, while in the $H\to Z\gamma$ channel the corrections for $s_{\beta-\alpha}\neq 1$ likewise cross zero and turn positive. Analytic expressions for the one-loop amplitudes in the alignment limit have been provided, taking $H\to\gamma\gamma$ as an example. Owing to $CP$ conservation, the intermediate $t\bar t$ pair is produced in the ${}^{3}P_{0}$ state, for which Coulomb effects are strongly suppressed and the threshold behaviour remains perturbative. The threshold expansion of the complete two-loop amplitude confirms that the QCD correction stays finite, with an imaginary part proportional to $\beta_t^2$ and a finite real part at threshold, in agreement with the expected $P$-wave behaviour. These results demonstrate that the fixed-order prediction provides a stable description of the overall size and threshold behaviour of the QCD corrections for the two processes, while any higher-order threshold effects are expected to introduce only finite quantitative modifications. The residual perturbative uncertainty, estimated by varying the renormalisation scale over $\mu\in[m_H/2,\,2m_H]$, is bounded by approximately $3.8\%$ across the viable parameter space, well below the typical size of the corrections themselves. A phenomenological analysis comparing $\sigma(gg\to H)\times\mathrm{Br}(H\to V\gamma)$ with ATLAS 95\% CL upper limits reveals that the $H\to\gamma\gamma$ predictions at NLO already reach approximately 83\% of the observed limit near the $t\bar{t}$ threshold, whereas the $H\to Z\gamma$ channel remains roughly 30 times below the current sensitivity. Should the experimental sensitivity approach the current exclusion limit with the increased luminosity at the HL-LHC, the $\mathcal{O}(15\%)$ NLO corrections may become relevant for the interpretation of the exclusion boundary in the $(m_H,\,\tan\beta)$ plane. These results furnish a comprehensive characterisation of the loop-induced decay properties of the heavy neutral Higgs boson in the Type-I THDM and may serve as valuable input for forthcoming experimental searches and precision measurements at hadron colliders.
	
	\appendix
	\section{Analytic expressions for the $H\to V\gamma$ amplitudes \label{sec:LOexpress}}
	
	In this appendix we collect the analytic expressions entering the calculation of $H\to V\gamma$ ($V=Z,\gamma$) within the Type-I THDM. The kinematic variables are
	\begin{equation}
		\tau_i \equiv \frac{m_H^2}{4m_i^2}\,,\qquad
		\lambda_i^V \equiv \frac{m_V^2}{4m_i^2}.
		\label{eq:apptau}
	\end{equation}
	
	The complete one-loop amplitude for $H\to V\gamma$ takes the compact form
	\begin{equation}
		\begin{aligned}
			A_4^{\text{(1-loop)}}(H\!\to\!V\gamma)=\frac{\alpha}{2\pi v}\Big[
			&\hat{g}_{V}^{t}N_c\,Q_t^2\,\,\kappa_{u,H}\,
			{\mathcal{A}}_{1/2}(\tau_t,\lambda^{V}_t)
			+\hat{g}_{V}^{H^{\pm}}\,\lambda_{\text{eff}}\,
			{\mathcal{A}}_{0}(\tau_{H^\pm},\lambda^{V}_{H^\pm})\\
			&+c_{\beta-\alpha}\,\hat{g}^{W}_{V}{\mathcal{A}}_{1}(\tau_W,\lambda^{V}_W)
			\Big]\,,
		\end{aligned}
		\label{eq:ampVgamma}
	\end{equation}
	
	where the coefficients are
	\begin{equation}
		\label{eq:couplings}
		\begin{aligned}
			N_c &= 3\,,\\[3pt]
			\hat{g}_{\gamma}^{t,H^{\pm}} &= 1\,,\qquad
			\hat{g}_{Z}^{t} = -\frac{1/2-2Q_t s_W^2}{2Q_t s_W c_W}
			\,,\\[3pt]
			\hat{g}_{Z}^{H^{\pm}} &= \frac{1/2-s_W^2}{s_W c_W}\,,\qquad
			\hat{g}^{W}_{V} = -\frac{c_W}{s_W}\,.
		\end{aligned}
	\end{equation}
	
	and the effective trilinear Higgs--charged-Higgs coupling is
	\begin{equation}
		\lambda_{\text{eff}}\equiv i\frac{v\,g_{HH^{+}H^{-}}}{2m^2_{H^\pm}}\,.
	\end{equation}
	
	The scalar form factors for spin-$\tfrac12$, spin-$0$, and spin-$1$ intermediate states are expressed in terms of the two basic integrals $I_1$ and $I_2$ as
	\begin{equation}
		\begin{aligned}\label{eq:mathcalA}
			\mathcal{A}_{1/2}(\tau,\lambda) &=\;4\big[I_2(\tau,\lambda)-I_1(\tau,\lambda)\big]\,,\\[5pt]
			\mathcal{A}_{0}(\tau,\lambda)   &=\;2I_1(\tau,\lambda)\,,\\[5pt]
			\mathcal{A}_{1}(\tau,\lambda)   &=\;\big[4(\tau-\lambda-2\tau\lambda)+6\big]\,I_1(\tau,\lambda)
			-16(1-\lambda)\,I_2(\tau,\lambda)\,.
		\end{aligned}	
	\end{equation}
	The integrals $I_1$ and $I_2$ are defined as
	\begin{equation}\label{eq:I1I2}
		\begin{aligned}
			I_1(\tau,\lambda)
			&= \frac{-1}{2(\tau-\lambda)}
			+ \frac{f(\tau)-f(\lambda)
				+ 2\lambda\big[g(\tau)-g(\lambda)\big]}
			{2(\tau-\lambda)^2}\,,\\[6pt]
			I_2(\tau,\lambda)
			&= \frac{f(\tau)-f(\lambda)}{2(\tau-\lambda)}\,,
		\end{aligned}
	\end{equation}
	where the primary functions $f(\tau)$ and $g(\tau)$ are
	\begin{equation}
		f(\tau) = \begin{cases}
			\arcsin^2\!\sqrt{\tau}\,, & \tau \le 1\,,\\[4pt]
			-\dfrac{1}{4}\!\left[\ln\dfrac{1+\sqrt{1-1/\tau}}{1-\sqrt{1-1/\tau}} - i\pi\right]^2, & \tau > 1\,,
		\end{cases}
		\qquad
		g(\tau)=\begin{cases}
			\sqrt{\tau^{-1}-1}\,\arcsin\sqrt{\tau}\,, & \tau\le1\,,\\[4pt]
			\tfrac12\sqrt{1-\tau^{-1}}\Big[\ln\dfrac{1+\sqrt{1-\tau^{-1}}}{1-\sqrt{1-\tau^{-1}}}-i\pi\Big], & \tau>1\,.
		\end{cases}
		\label{eq:ftau}
	\end{equation}
	
	For completeness, we also list the two-loop form factor $\mathcal{A}_{1/2}^{\text{(2-loop)}}$ for the $H\to\gamma\gamma$ channel, which corresponds to the $\mathcal{A}_{1/2}^{\text{(2-loop)}}$ term in Eq.~(\ref{eq:A4gammagamma_align}):
	\begin{align}
		\mathcal{A}_{1/2}^{\text{(2-loop)}}={}&
		-\frac{\theta(1+\theta+\theta^2+\theta^3)}{(1-\theta)^5}
		\Bigg[
		108\,\operatorname{Li}_4(\theta)
		+144\,\operatorname{Li}_4(-\theta)
		-64\,\operatorname{Li}_3(\theta)\ln\theta
		\nonumber\\
		&
		-64\,\operatorname{Li}_3(-\theta)\ln\theta
		+14\,\operatorname{Li}_2(\theta)\ln^2\theta
		+8\,\operatorname{Li}_2(-\theta)\ln^2\theta
		+\frac{1}{12}\ln^4\theta
		\nonumber\\
		&
		+4\zeta_2\ln^2\theta
		+16\zeta_3\ln\theta
		+18\zeta_4
		\Bigg]
		\nonumber\\[1ex]
		&
		+\frac{\theta(1+\theta)^2}{(1-\theta)^4}
		\Big[
		-32\,\operatorname{Li}_3(-\theta)
		+16\,\operatorname{Li}_2(-\theta)\ln\theta
		-4\zeta_2\ln\theta
		\Big]
		\nonumber\\
		&
		-\frac{4\theta(7-2\theta+7\theta^2)}{(1-\theta)^4}
		\operatorname{Li}_3(\theta)
		+\frac{8\theta(3-2\theta+3\theta^2)}{(1-\theta)^4}
		\operatorname{Li}_2(\theta)\ln\theta
		\nonumber\\
		&
		+\frac{2\theta(5-6\theta+5\theta^2)}{(1-\theta)^4}
		\ln(1-\theta)\ln^2\theta
		+\frac{\theta(3+25\theta-7\theta^2+3\theta^3)}
		{3(1-\theta)^5}
		\ln^3\theta
		\nonumber\\
		&
		+\frac{4\theta(1-14\theta+\theta^2)}{(1-\theta)^4}\zeta_3
		+\frac{12\theta^2}{(1-\theta)^4}\ln^2\theta
		-\frac{12\theta(1+\theta)}{(1-\theta)^3}\ln\theta
		-\frac{20\theta}{(1-\theta)^2},
		\label{eq:2loopA}
	\end{align}
	with
	\begin{equation}
		\theta\equiv \dfrac{\sqrt{1-\tau^{-1}}-1}{\sqrt{1-\tau^{-1}}+1}\,.
	\end{equation}
	
	For the $H\to Z\gamma$ channel, the two-loop form factor involves 28 master integrals depending on three mass scales, and is expressed through logarithms, the polylogarithms $\mathrm{Li}_n$ ($n=2,3,4$), and the two-variable function $\mathrm{Li}_{2,2}$. We evaluate these master integrals with \texttt{AMFlow} and find agreement with Ref.~\cite{Bonciani:2015eua}. The full analytic expression is too lengthy to present here.
	
	\clearpage
	
	\bibliographystyle{apsrev4-2}
	\bibliography{bibliography}
	
\end{document}